\documentclass[11pt,a4paper]{article}
\usepackage{adjustbox}
\usepackage{jinstpub}
\usepackage{lineno}
\usepackage{orcidlink}
\usepackage{color}
\usepackage{siunitx}
\usepackage{subcaption}
\usepackage{url}

\DeclareSIUnit\bar{bar}

\newcommand{\iselSystem}{isel-system}

\newcommand{\figsubref}[1]{(\protect\subref{#1})}
\newcommand{\figref}[1]{Figure~\ref{#1}}
\newcommand{\figrefbra}[1]{Fig.~\ref{#1}}
\newcommand{\Figref}[1]{Figure~\ref{#1}}

\newcommand{\secref}[1]{Section~\ref{#1}}
\newcommand{\secrefbra}[1]{Sec.~\ref{#1}}

\newcommand{\eqnref}[1]{Eqn.~\eqref{#1}}

\title{GRANITE: High-Resolution Imaging and Electrical Qualification of Large-Area TPC Electrodes}

\author[a,1]{Shumit~Mitra\,\orcidlink{0009-0008-0610-6152},\note{Corresponding author}}
\emailAdd{putumittir@googlemail.com}
\author[a]{Alexander~Deisting,\,\orcidlink{0000-0001-5372-9944}}
\author[a]{Jan~Lommler\,\orcidlink{0009-0004-9049-2199},}
\author[a,b]{Uwe~Oberlack\,\orcidlink{0000-0001-8160-5498},}
\author[b]{Fabian~Piermaier\,\orcidlink{0000-0002-3863-3394},}
\author[b]{Quirin~Weitzel\,\orcidlink{0000-0002-3073-8642},}
\author[a,2]{and Daniel~Wenz\,\orcidlink{0009-0004-5242-3571}\note{Now at: Universität Münster, Institut für Kernphysik, Münster (DE)}} 
\affiliation[a]{Institut für Physik \& Exzellenzcluster PRISMA$^{+}$, Johannes Gutenberg-Universität Mainz, Staudingerweg 7, 55128 Mainz, Germany}
\affiliation[b]{Detektorlabor, Exzellenzcluster PRISMA$^{+}$, Johannes Gutenberg-Universität Mainz, Staudingerweg 9, 55128 Mainz, Germany}

\abstract{Next-generation dual-phase time projection chambers (TPCs) for rare event searches will require large-scale, high-precision electrodes. To meet the stringent requirements for high-voltage performance of such an experiment, we have developed a scanning setup for comprehensive electrode quality assurance. The system is built around the GRANITE (Granular Robotic Assay for Novel Integrated TPC Electrodes) facility: a gantry robot on top of a $\SI{2.5}{\meter}\times\SI{1.8}{\meter}$ granite table, equipped with a suite of non-contact metrology devices. 
  We developed a coaxial wire scanning head to measure and correlate localized high-voltage discharge currents in air with high-resolution surface images. We find that the identified discharge 'hotspots' are transient and show no significant correlation with static visual features. Next, we established a quantitative relationship between artificially induced abrasive surface damage on the wires and a reduction in the discharge inception voltage. This work provides a novel non-invasive tool for qualifying wires dedicated for use in electrodes for future low-background experiments.}


\arxivnumber{2511.11401}

\begin{document}

\maketitle

\section{Introduction}
\label{sec:introduction}

Liquid xenon (LXe) dual-phase time projection chambers (TPCs) have set the most stringent limits on weakly interacting massive particle (WIMP) dark matter \cite{PhysRevLett.131.041003,4dyc-z8zf,PhysRevLett.134.011805} for WIMP masses $\gtrsim\!\!\SI{5}{\giga\electronvolt\per c^2}$. These detectors result from extensive R\&D efforts aimed at increasing target mass and reducing intrinsic backgrounds. Planning has begun for the next-generation xenon observatory, covering a broad rare-event programme including coherent elastic neutrino–nucleus scattering (CE$\nu$NS), first indications of which have been reported \cite{PhysRevLett.133.191002,PhysRevLett.133.191001}.\\
{\indent}The XLZD collaboration proposes a dual-phase TPC with $\sim\!\!\SI{3}{\meter}$ diameter and height \cite{XLZD:2024nsu}, based on the success of the XENONnT \cite{XENON:2024wpa} and LZ \cite{AKERIB2020163047} experiments. These, together with PandaX-4T \cite{PhysRevLett.127.261802}, employ electrodes of $\SI{1}{\meter}$–$\SI{1.5}{\meter}$ diameter, so XLZD represents more than a twofold scale-up. The TPC has three main electrodes: cathode, gate, and anode. The gate and anode enclose the liquid–gas interface, while the cathode sits at the bottom. A particle interaction produces primary scintillation light (S1) and ionization electrons, which drift in the field between cathode and gate toward the top. A strong field between gate and anode extracts these electrons across the interface and imparts them with enough energy to generate electroluminescence light (S2) in the xenon gas. Both S1 and S2 are detected by light sensor arrays at the top and bottom of the TPC.\\
{\indent}Microscopic asperities or surface defects on electrodes can trigger field emission, especially under high external fields. Emitted electrons can mimic low-energy events, contribute to accidental (coincidence) backgrounds, or in severe cases, cause breakdown between electrodes. Reported emission thresholds range from $\sim\!\!\SI{1E+4}{\volt\per\centi\meter}$ \cite{TOMAS201849} to $\sim\!\!\SI{1E+6}{\volt\per\centi\meter}$ \cite{raizer1997gas}, the former observed in dual-phase xenon where general surface treatment rather than single protrusions were found to dictate emission behaviour. In addition, the Malter effect \cite{PhysRev.50.48} can occur when a cathodic surface (e.g.\ the gate or cathode) develops a thin insulating layer, leading to localised positive charge build-up and consequent secondary electron emission. This mechanism, well known in gaseous detectors \cite{Rolandi:2008ujz,raizer1997gas}, has recently been discussed in the LXe context \cite{Vavra:2025fif}.\\
{\indent}The fabrication of dual-phase TPC electrodes thus demands rigorous inspection and preparation of wire materials. Typical procedures include cleaning (e.g.\ ultrasonic baths with alkaline soaps), rinsing, and passivation with dilute citric or nitric acid to form a stable oxide layer \cite{LINEHAN2022165955,XENON:2024wpa}. Afterward, wires are stretched and mounted on frames \cite{LINEHAN2022165955,Stifter_2020}. Inspection may rely on optical imaging, high voltage (HV) performance testing, or both. Optical inspection, even when automated, is time-consuming \cite{Wenz:2023qzq}. Following initial optical and HV studies \cite{Wenz:2023qzq,Mitra:2023,Deisting2025}, we developed metrology tools that identify defects through monitoring dark-regime discharge currents.\\
{\indent}Testing electrodes of XLZD scale in cryogenic LXe would be time-intensive and costly. Hence, preliminary qualification in air or inexpensive gases is essential before cryogenic validation, \textit{e.g.} in  \cite{brown2024pancake}. We therefore established the quality-assurance procedures presented here. These are an extension of the set-up called ``GRANITE'', short for ``Granular Robotic Assay for Novel Integrated TPC Electrodes'', which is covered in detail in \cite{Deisting2025}. That paper introduces the set-up, its ability for mechanical characterisation of electrodes, and demonstrates an optical scan with a high resolution camera of the XENON1T \cite{XENON:2017lvq} cathode. GRANITE can scan electrodes over areas exceeding \SI{2}{\meter\squared}, with methods readily scalable to larger assemblies. During the optical scans in \cite{Deisting2025}, it became evident that high resolution images alone are not sufficient to draw conclusions about the nature of anomalies on the wire surface. Therefore, the studies in this paper were initiated.\\
{\indent}In \secref{sec:setup} we describe the experimental setup, followed by the defect-detection techniques combining high-resolution imaging and dark-discharge current monitoring (\secrefbra{sec:highres-scans}). The HV scanning tool and relevant discharge phenomenology are introduced, and results on correlations between visual features and current emission are discussed. Sections \ref{sec:highres-scans:subsec:high-voltage-wire-scanning}–\ref{sec:highres-scans:subsec:visual_correlation} present the measurements and derived sensitivity thresholds for defect-induced emission. Wires showing higher-than-nominal discharge currents for a given field can thus be rejected from electrode construction. Finally, we conclude in \secref{sec:conclusions:outlook} with an outlook on integrating these methods into large-scale electrode production.

\section[The PRISMA+ Electrode Assay Setup]{The PRISMA$^{+}$ Electrode Assay Setup}
\label{sec:setup}

\begin{figure}[htbp]
  \centering
  \subfloat[]{
    \label{sec:setup:fig:granite_table:schematic}
    \includegraphics[height=0.24\textheight]{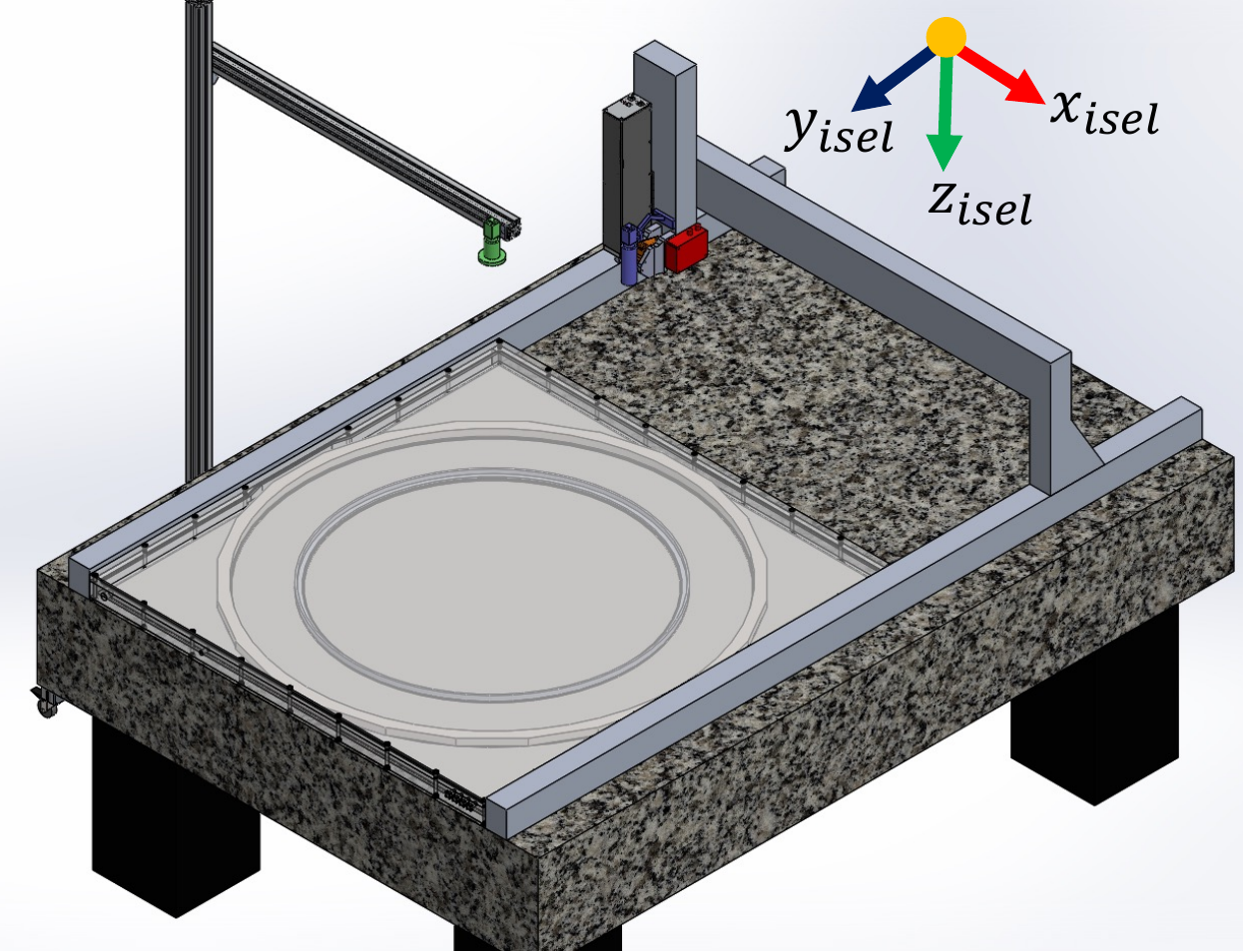}}\hspace{0.2cm}%
  \subfloat[]{
    \label{sec:setup:fig:granite_table:metrology}
    \includegraphics[height=0.24\textheight,trim= 20 0 100 100, clip=true]{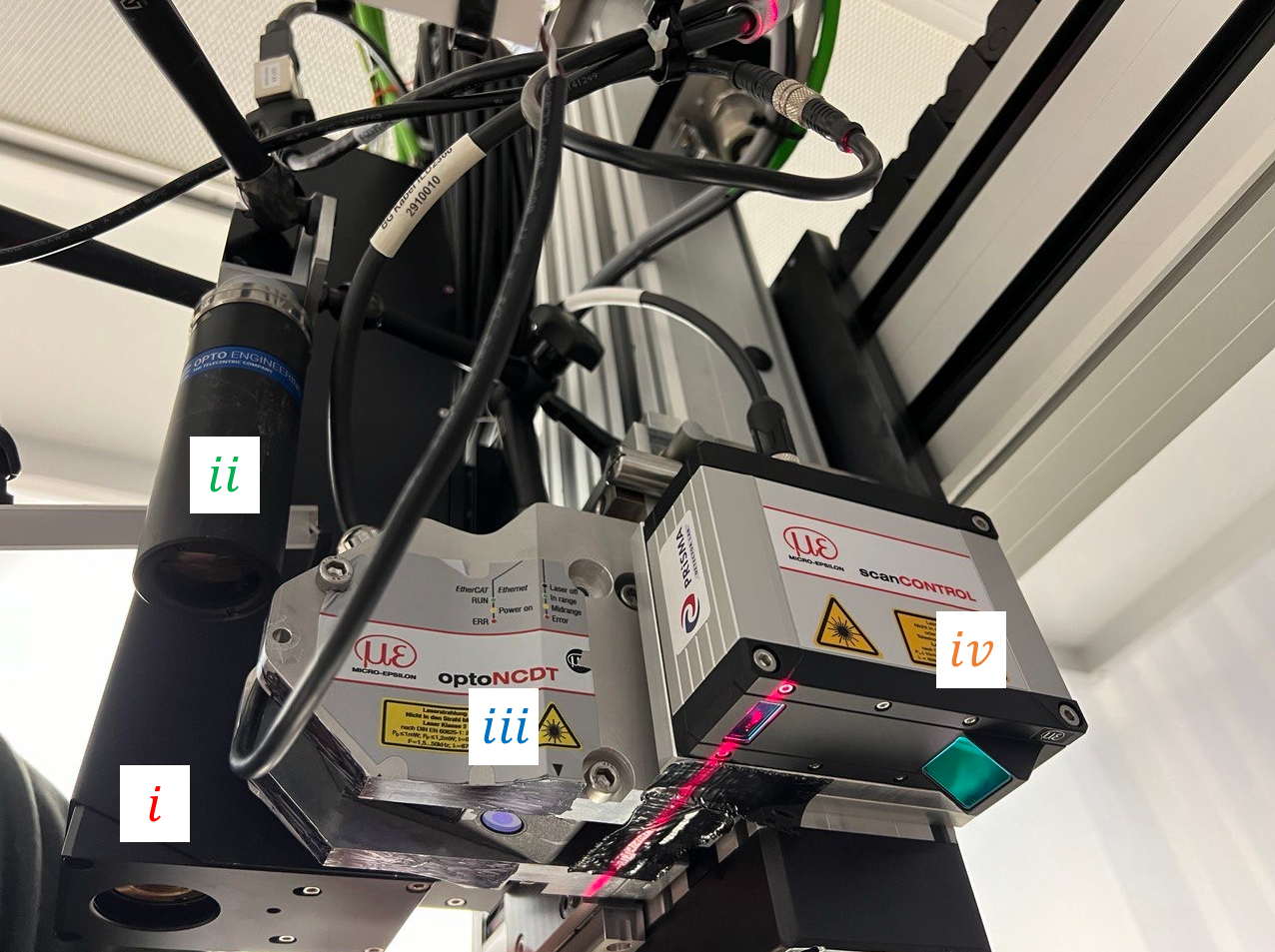}}\\[1em]

  \begin{minipage}[b]{0.3\linewidth}
    \centering
    \subfloat[]{
      \label{sec:setup:fig:opticalSensors:confocal-microscope}
      \includegraphics[width=\linewidth]{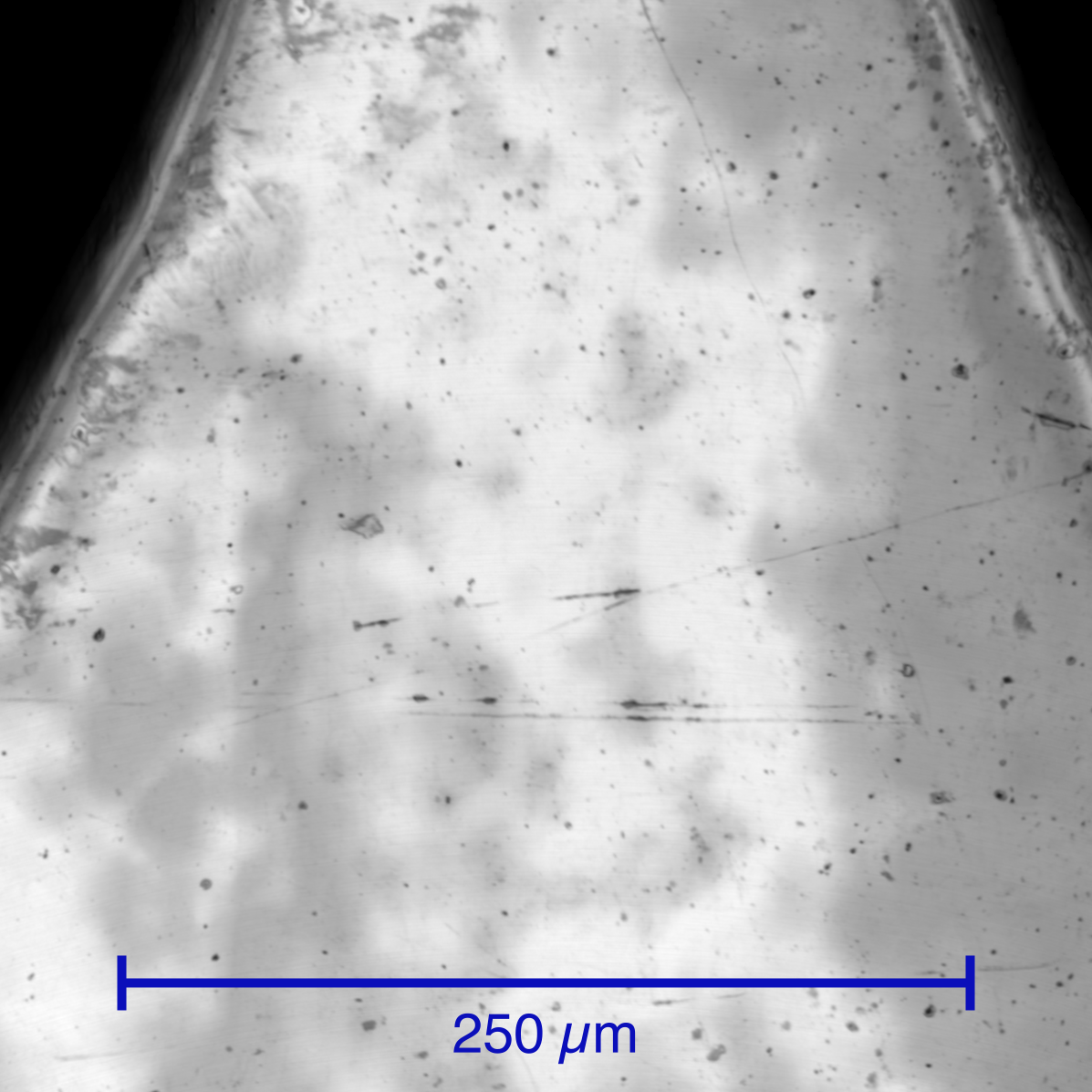}}\\[-0.3em]
    \subfloat[]{
      \label{sec:setup:fig:opticalSensors:high-res-cam}
      \includegraphics[width=\linewidth]{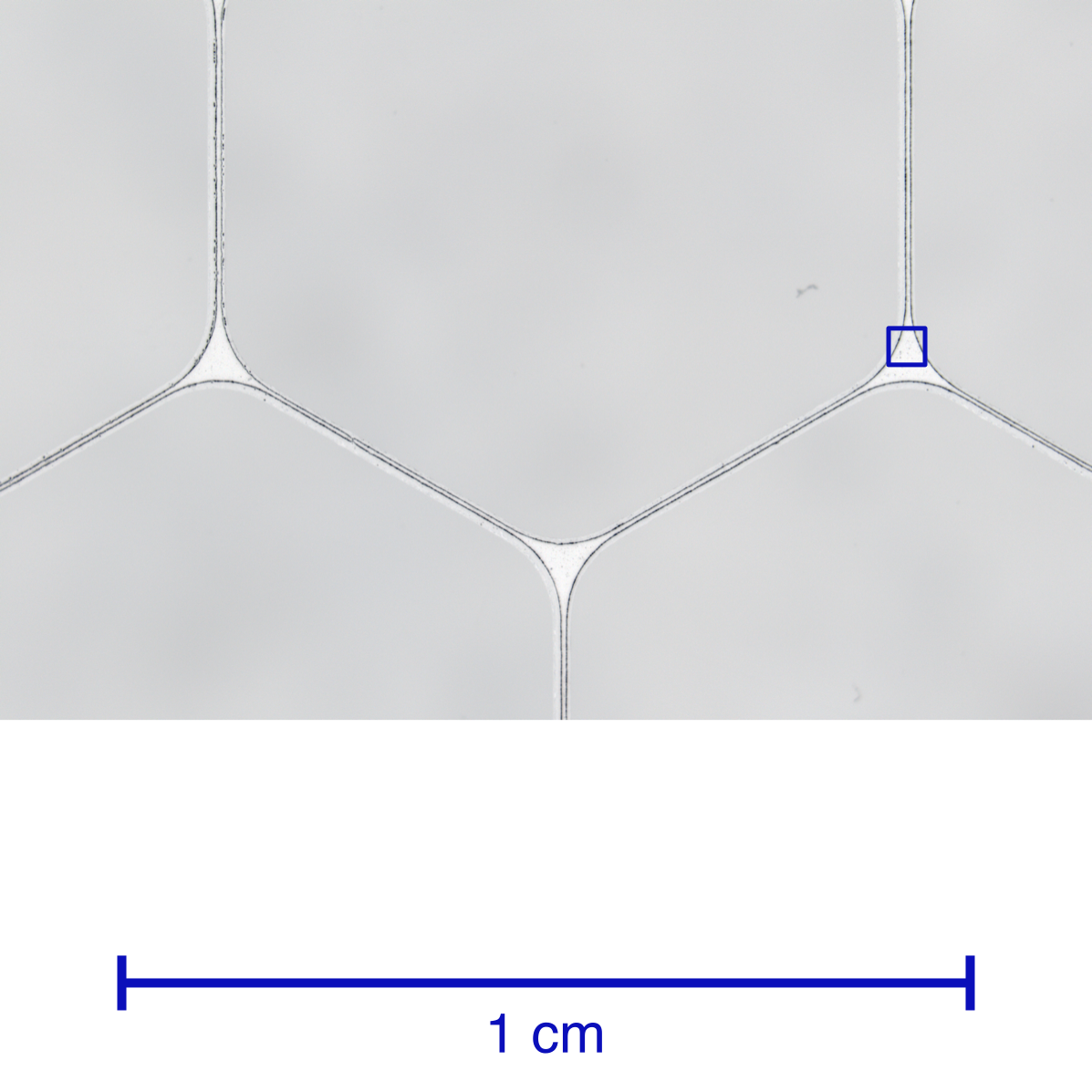}}
  \end{minipage}\hspace{0.2cm}
  \subfloat[]{
    \label{sec:setup:fig:granite_table:photo}
    \adjustbox{valign=b}{\includegraphics[height=0.40674\textheight]{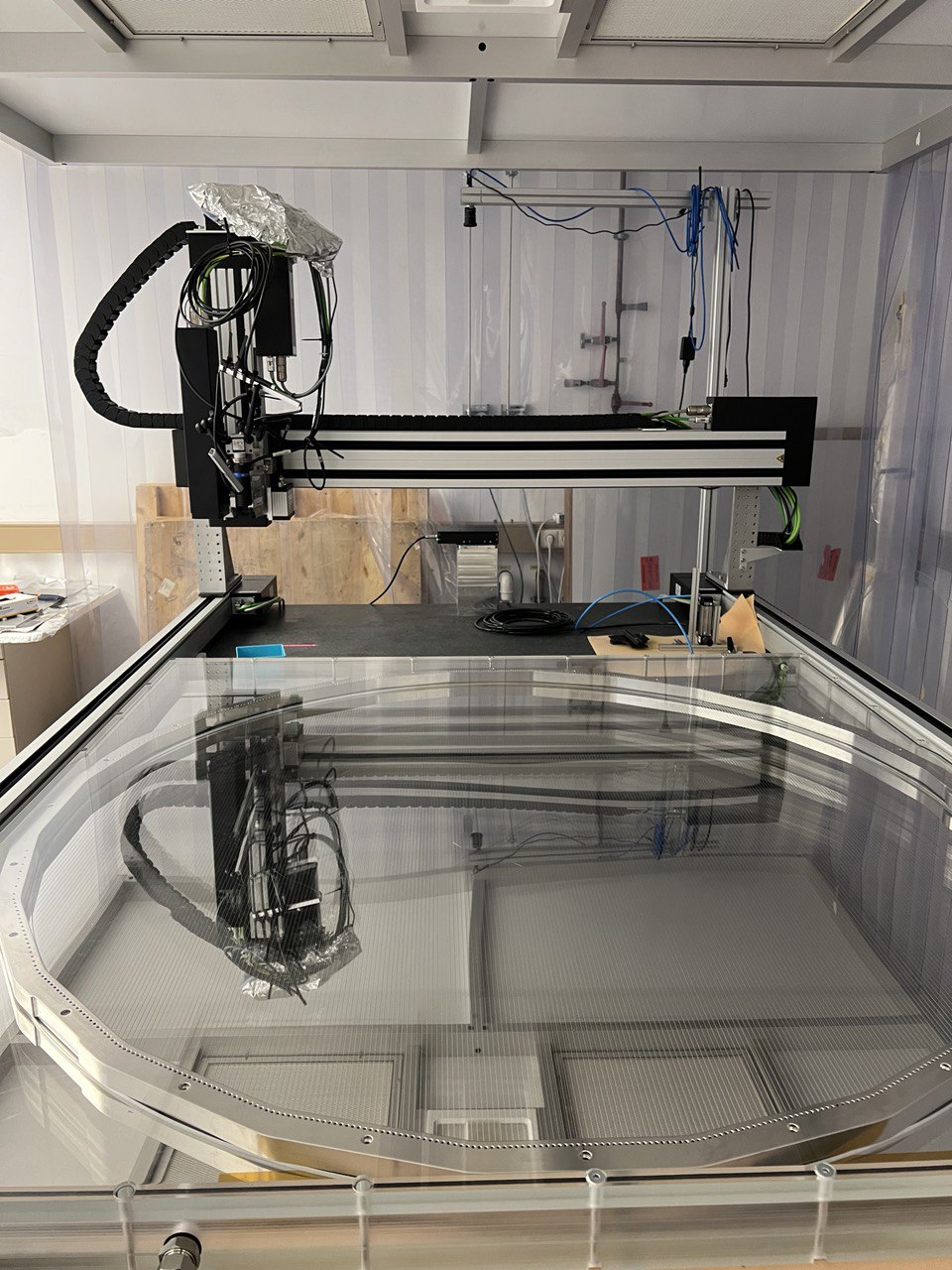}}}

  \caption{\label{sec:setup:fig:granite_table}
  \figsubref{sec:setup:fig:granite_table:schematic} Schematic of the gantry setup on the $2.5 \times \SI{1.8}{\meter\squared}$ granite table with coordinate system. 
  \figsubref{sec:setup:fig:granite_table:metrology} The different metrology components from left to right: Confocal microscope (\textcolor{red}{$i$}) and -- slightly higher -- the high resolution (industrial) camera (\textcolor{green!45!black}{$ii$}), the laser distance sensor (\textcolor{blue}{$iii$}), and the profile laser scanner (\textcolor{orange}{$iv$}). 
  \figsubref{sec:setup:fig:opticalSensors:confocal-microscope} Illustration of the resolution of the \textit{confocal microscope} and a $\times50$ lens, and 
  \figsubref{sec:setup:fig:opticalSensors:high-res-cam} the \textit{high resolution industrial camera}. The region marked in blue shows the size of the confocal microscope image.
  \figsubref{sec:setup:fig:granite_table:photo} Photo of the setup: In the foreground an acrylic glass box with a $\sim\!\!\SI{1.4}{\meter}$ diameter electrode can be seen, whilst the arm with the metrology components described in the text is visible on the far, left side of the table.}
\end{figure}
The setup, shown in \figref{sec:setup:fig:granite_table}, is built on a granite table, providing a smooth and level surface of roughly $2.5 \times \SI{1.8}{\meter\squared}$, as well as vibration damping due to its mass. A laminar flow unit housing the granite table is used to keep the lab environment clean. The whole setup is located in a temperature controlled lab, and temperature, together with relative humidity, dew-point and atmospheric pressure are recorded. A gantry robot is mounted on top of the table (``\iselSystem{}'') built from parts provided by iselGermany, holding multiple optical measurement devices: 
(\textit{i}) a confocal microscope (\textit{NanoFocus µsurf 350 HDR F}, example image is shown in \figref{sec:setup:fig:opticalSensors:confocal-microscope}), 
(\textit{ii}) an industrial camera \cite{baslerweb:acA4600-7gc} (\textit{Basler acA4600-7gc}) with a telecentric lens \cite{opto-e:TC23016} (field depth \SI{1.5}{\milli\meter}, working distance \SI{43.1}{\milli\meter}, example image \figref{sec:setup:fig:opticalSensors:high-res-cam}), 
(\textit{iii}) a laser distance sensor \cite{muepsilon:optoNCDT2300} (\textit{optoNCDT 2300-20}), and 
(\textit{iv}) a profile laser scanner \cite{muepsilon:scanCONTROL3000} (\textit{scanCONTROL 3000}). Each of these devices can be used for a different measurement:
\begin{itemize}
  \setlength\itemsep{-0.1em}
  \item 3D surface imaging of the wire with sub-micron precision (\textit{i}) 
  \item High quality pictures of the wire surface (\textit{ii}) 
  \item Sagging measurement of a single wire or multiple wires simultaneously, calibration measurements of the table and gantry (\textit{iii}), (\textit{iv}) 
  \item Measurements of distance oscillations (\textit{iii}) 
\end{itemize}
{\indent}The measurement devices can be moved with the gantry robot enabling scans of surfaces up to an area of about $2.0 \times \SI{1.4}{\meter\squared}$. 
Laser distance sensor and \iselSystem{} are identical to \cite{8069756}. A repeatability of $\pm\SI{20}{\micro\meter}$ for repeatedly moving the gantry to a given position anywhere on the table is quoted by the producer. The minimum step-size in direction of $x_\text{isel}$ and $y_\text{isel}$ ($z_\text{isel}$) is \SI{5}{\micro\meter} (\SI{2.5}{\micro\meter}). 
Custom \textsc{python} software controls the movement of the gantry robot, and the data acquisition of the camera and the laser distance sensors, as well as the CAEN power supplies.\\
{\indent}A custom acrylic glass enclosure has been made which facilitates the inspection of the wire electrodes under different electric fields in a controlled atmosphere (e.g. argon). Inside the enclosure is an ultra-high polished stainless steel plate for grounding. Electrodes lie on top of acrylic glass spacers. With this configuration, the electrodes can be examined at the desired electric fields. HV is supplied from different CAEN power supply modules \cite{CAEN:N1470,CAEN:DT1471ET}.

\section{High Resolution Scans of Wires}
\label{sec:highres-scans}

\begin{figure}
  \centering
    \subfloat[]{
      \includegraphics[height=0.18\textheight]{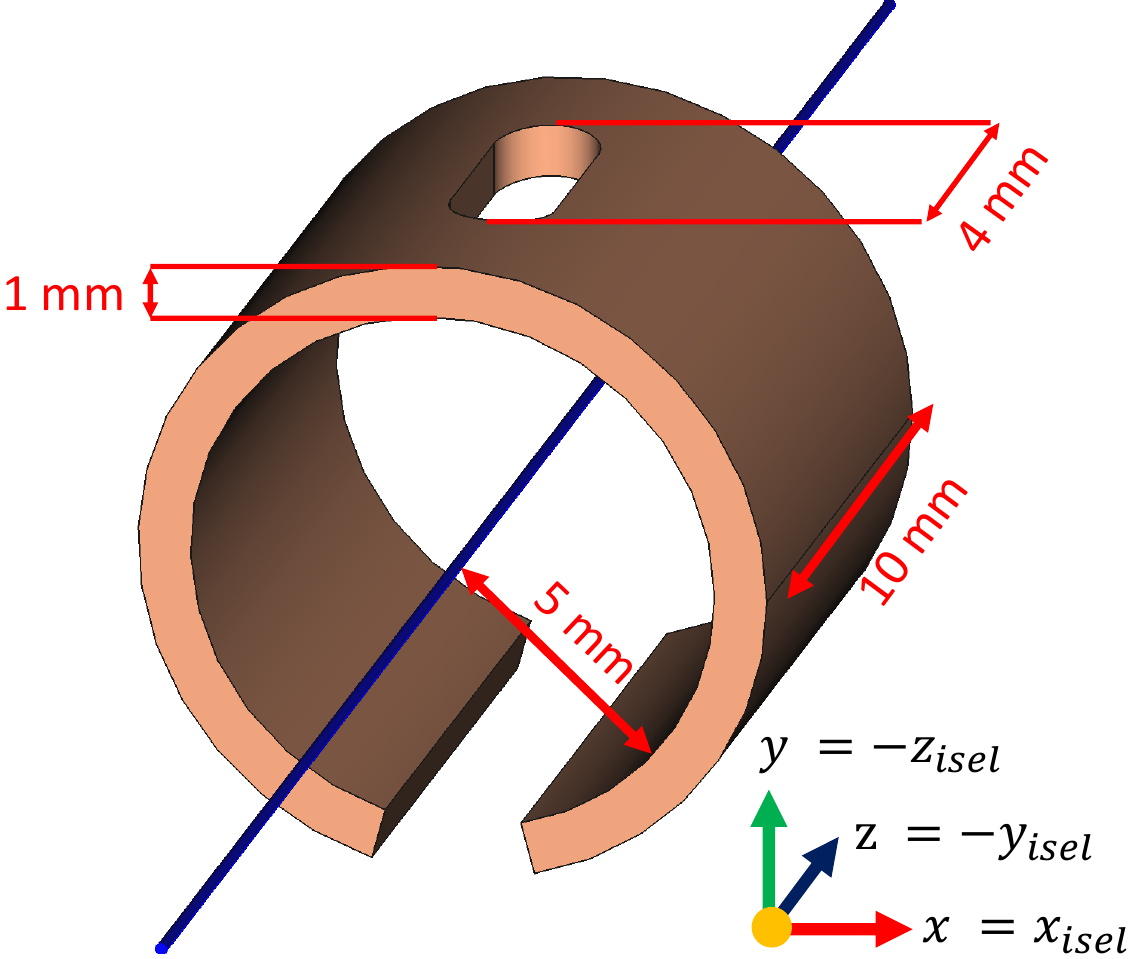}
      \label{fig:experimental_setup_coaxial_ground}}
    \subfloat[]{
      \includegraphics[height=0.18\textheight]{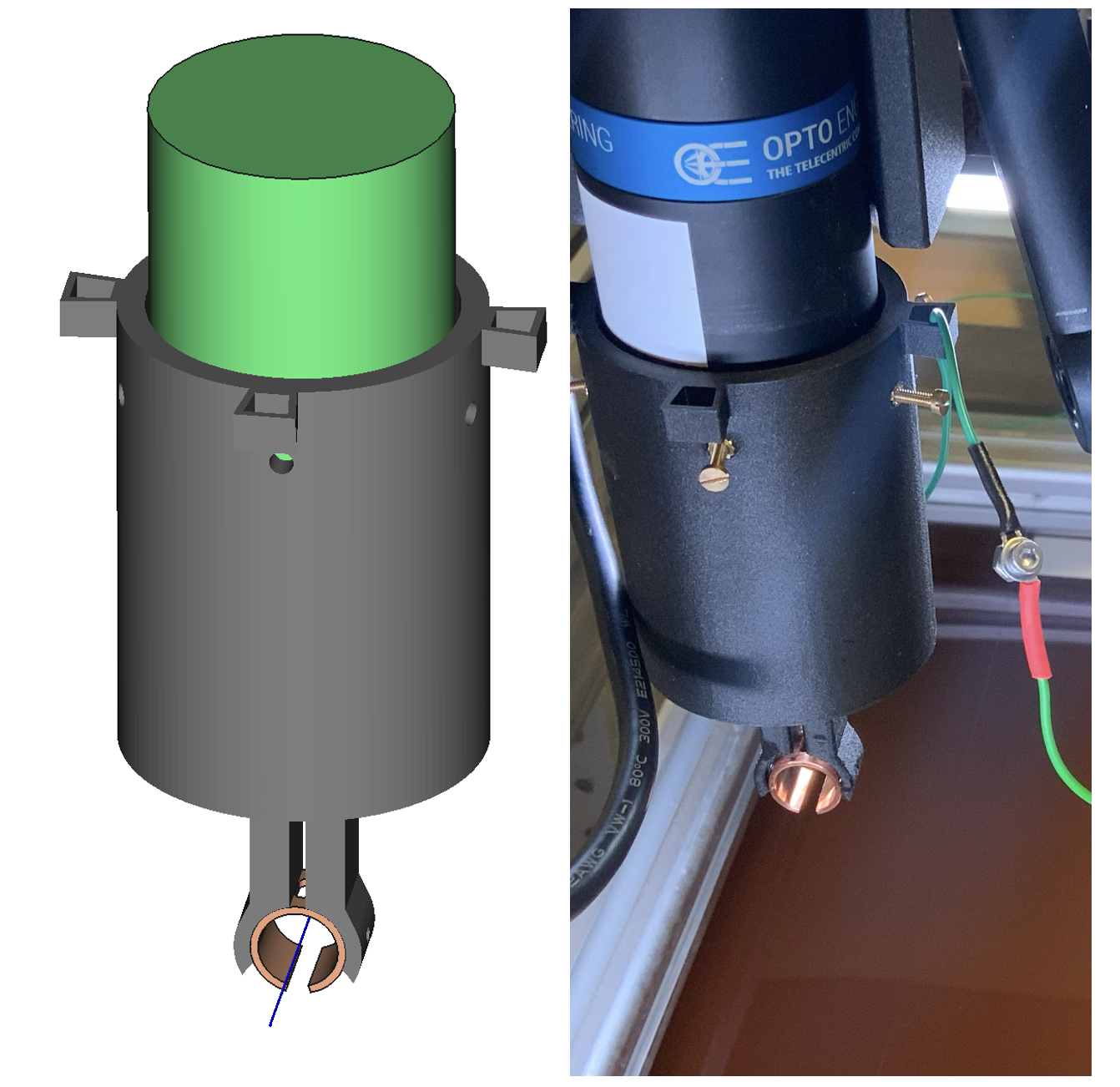}
      \label{fig:experimental_setup_scanning_head}}
    \subfloat[]{
      \includegraphics[height=0.18\textheight]{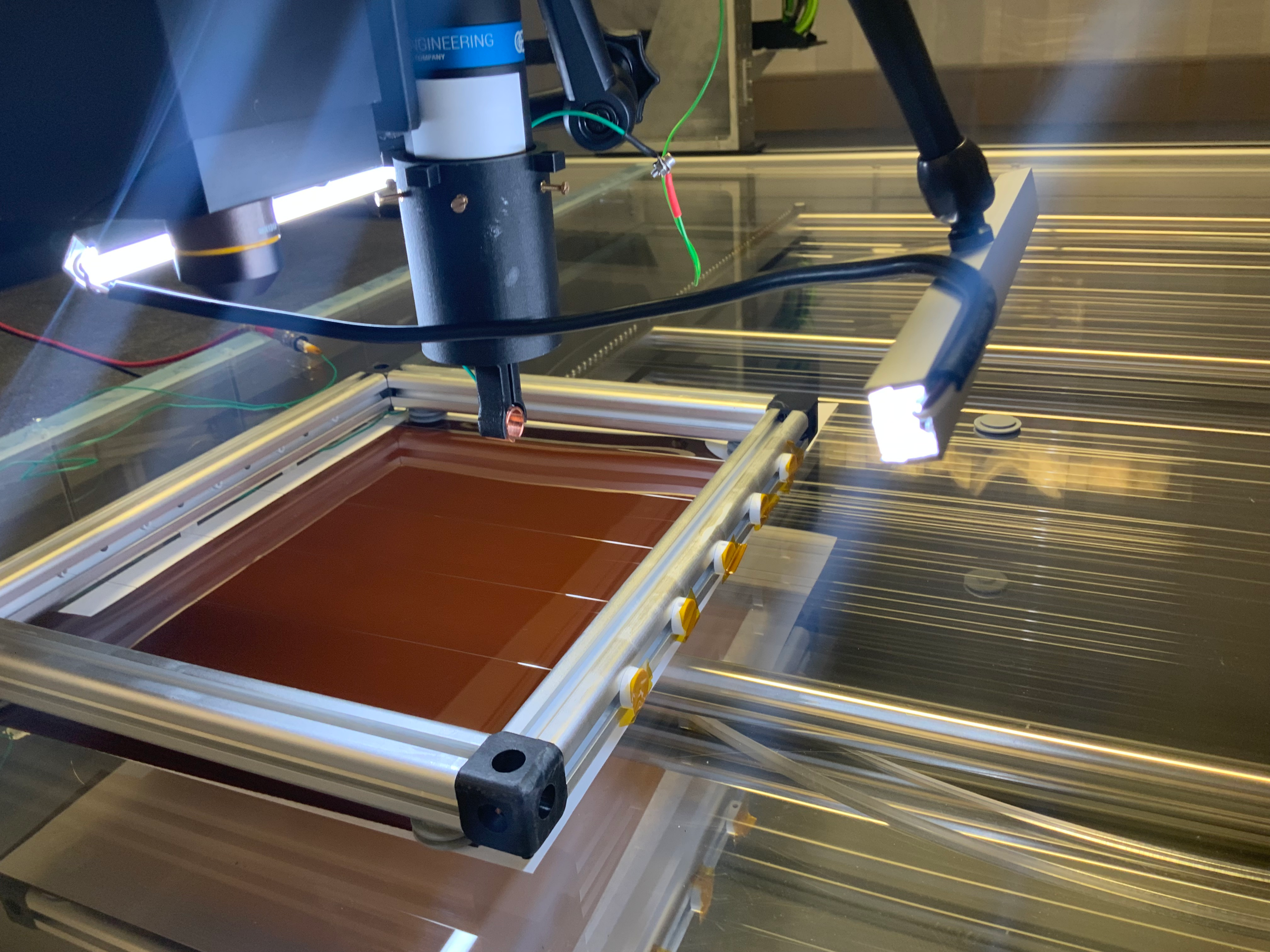}
      \label{fig:experimental_setup_5wire_frame}}\\[0.3cm]
    \subfloat[]{
    \includegraphics[width=0.75\columnwidth,angle=0]{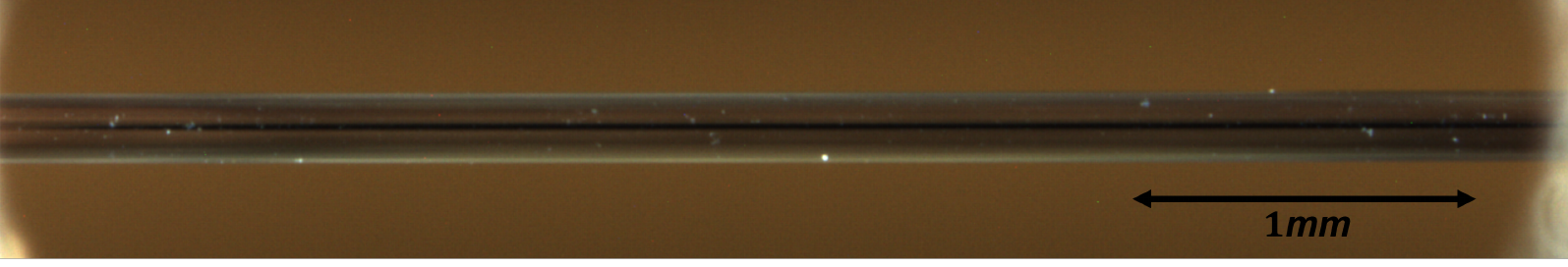}
      \label{fig:experimental_setup_camera_image}}
 \caption[Experimental setup for High Voltage Wire Scanning]{\label{fig:experimental_setup_hv_scanning}
      Experimental setup for wire scanning. The custom-designed scanning head is mounted on the telecentric lens of the \textit{Basler acA4600-7gc} camera, which is mounted on the gantry system. The cylindrical ground generates a nearly uniform field on the surface of the wire and includes a slit for the wire. 
      \figsubref{fig:experimental_setup_coaxial_ground} Zoomed-in view of the copper ground with the defined coordinate system (along with equalities to relate the axes to the \iselSystem{}'s coordinate system axes defined in \figref{sec:setup:fig:granite_table:schematic}): $y$ axis along the camera viewing direction, $z$ axis along the wire, and $x$ axis perpendicular to both. 
      \figsubref{fig:experimental_setup_scanning_head} 3D design and photo of the scanning head, showing the cylindrical ground and its camera attachment. 
      \figsubref{fig:experimental_setup_5wire_frame} The 5-wire frame used for HV wire scanning, optical scanning, and cleaning multiple wires. 
      \figsubref{fig:experimental_setup_camera_image} Close-up of the wire through the top slit in the coaxial configuration along with a \SI{1}{\milli\meter} scale.}
\end{figure}
Beyond meeting the stringent requirements for mechanical sagging and tension, the electrode wires must also exhibit exceptional high-voltage performance to prevent background-inducing discharges. In \cite{Deisting2025}, sagging and tension measurements using GRANITE have been discussed, as well as an optical scan of the XENON1T cathode using the \textit{Basler acA4600-7gc} camera, searching for surface anomalies. In these images a multitude of optical anomalies have been found on all wires of the cathode grid. This raises the question of which of these anomalies actually indicate a defect, which would limit later on the detector's performance. Therefore a study is needed, combining imaging methods with local HV tests along the wire. The remainder of this section describes this study, which is conducted on single wires. These are carefully selected and cleaned, to ensure that they are of a quality fit for dual-phase TPC of a dark matter (DM) experiment.

\subsection{High Voltage Scanning Setup and Measurement Procedure}
\label{sec:highres-scans:subsec:setup-measurement-procedure}

Stainless steel wires are examined for potential defect locations using high-resolution images of the \textit{Basler acA4600-7gc} camera, and current measurements obtained with a custom-designed scanning head, positioned coaxially to the wire. The scanning head consists of a cylindrical copper piece with dimensions shown in \figref{fig:experimental_setup_coaxial_ground}. It has a centred viewing hole for the camera on top and a \SI{3}{\milli\meter} slit at the bottom, allowing the head to move over a wire. The grounded copper piece is glued to a 3D-printed PLA holding structure, which is mounted to the \textit{Basler acA4600-7gc} camera’s telecentric lens' (\figrefbra{fig:experimental_setup_scanning_head}). This design ensures, that the centre of the copper cylinder is always in the focal plane. \Figref{fig:experimental_setup_camera_image} shows a cropped camera image taken through the viewing hole of the grounded electrode. The copper ground (scanning head) surrounds the wire coaxially. 
Preliminary studies on visible corona discharges from wires in argon and air are reported in \cite{Mitra:2023}. Five wires are fixed on a small $\SI{24}{\centi\meter}\times\SI{26}{\centi\meter}$ frame (\figrefbra{fig:experimental_setup_5wire_frame}). 
They were cut from a spool of California Fine Wire \cite{cfw:stainlesssteel} stainless steel wire, which is used in the XENONnT experiment. 
Before the first measurement, the loaded frame is cleaned in an ultrasonic bath with deionised water and an alkaline soap at $45\,^\circ\text{C}$. 
For the subsequent passivation, the wires are immersed in deionised water with an addition of \SI{7}{\%} citric acid at $45\,^\circ\text{C}$. The ultrasound is activated for \SI{10}{min} within the hour during which the wires rest inside the citric acid solution. 
Given previous experience \cite{flombardi2024}, the time with activated ultrasound is shorter and the citric acid fraction smaller than what has been used in the past. Otherwise, this cleaning procedure follows the general steps outlined in \secref{sec:introduction} and used for the works in \cite{LINEHAN2022165955,XENON:2024wpa}. After drying, the frame is immediately transferred to the measurement setup, where the laminar flow unit (\textit{cf}. \secref{sec:setup}) ensures environmental cleanliness. In addition, temperature, relative humidity, dew point, and atmospheric pressure are monitored.\\
\begin{figure}
  \centering
  \subfloat[]{
    \label{fig:sim_coax_field_singleXY}
    \includegraphics[width=0.45\columnwidth]{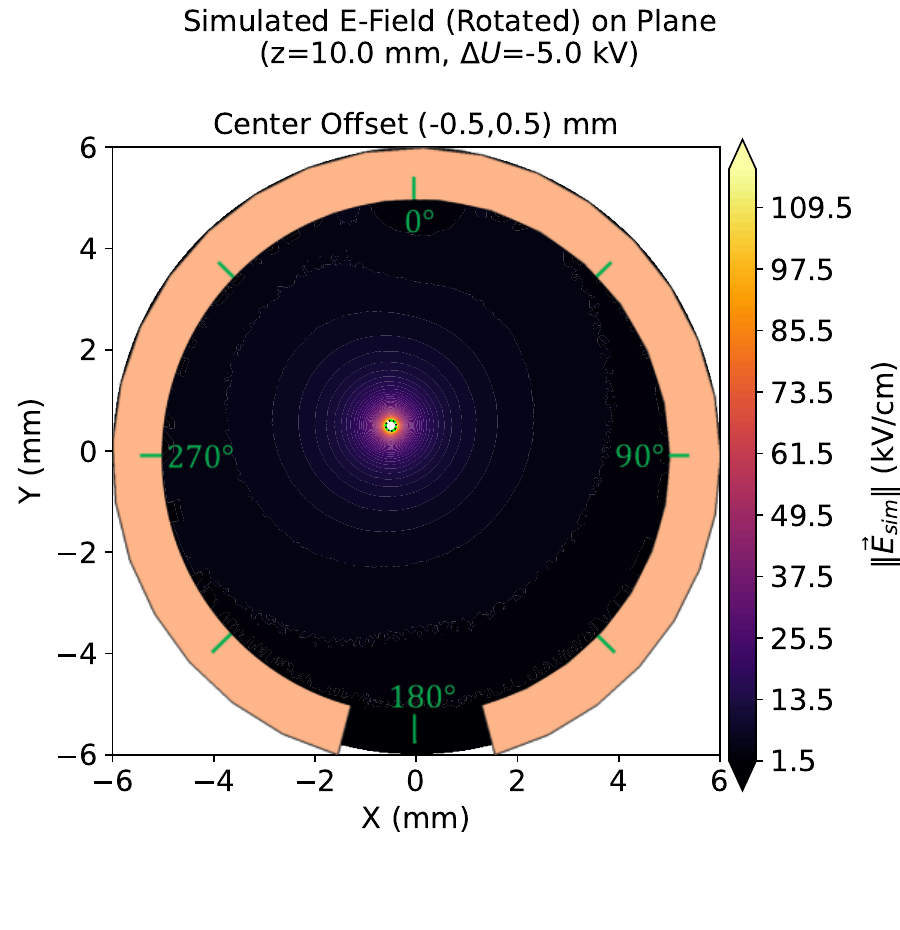}}
  \subfloat[]{
    \label{fig:sim_coax_field_res_singlexy}
    \includegraphics[width=0.45\columnwidth]{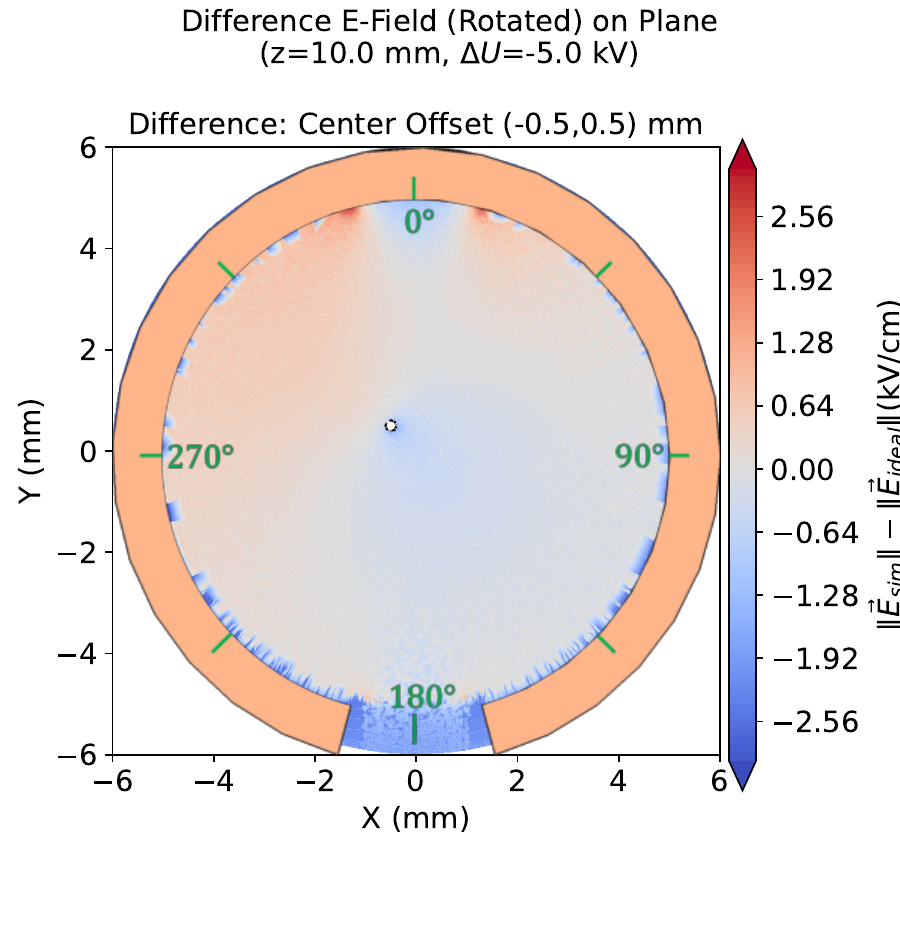}}\\[1em]

  \subfloat[]{
    \label{fig:sim_fitted_E_zTh}
    \includegraphics[width=0.45\columnwidth]{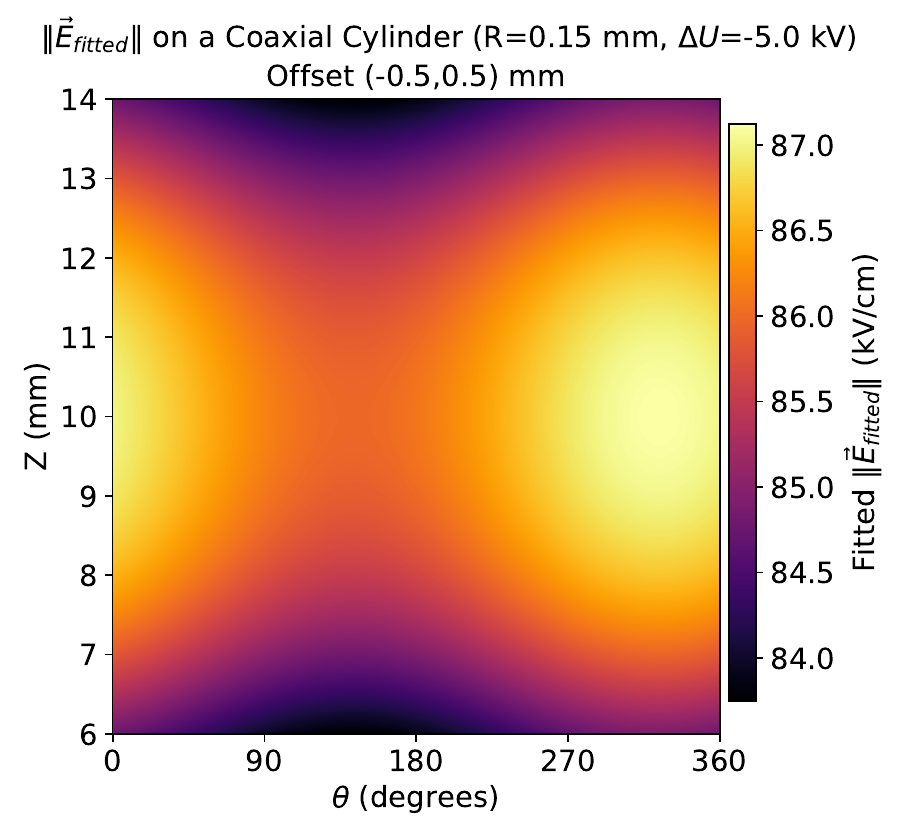}}
  \subfloat[]{
    \label{fig:sim_fitted_res_E_zTh}
    \includegraphics[width=0.45\columnwidth]{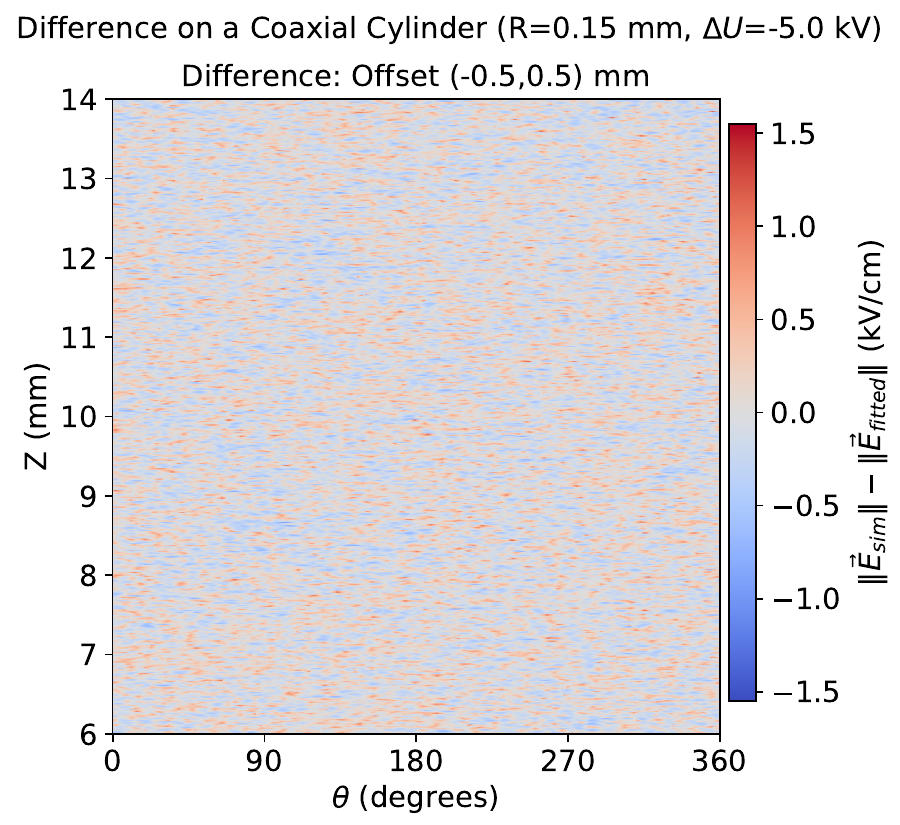}}
  \caption{\label{sec:FEM_simulations_z05_neg05_offset_single}
  Subplots \figsubref{fig:sim_coax_field_singleXY} and \figsubref{fig:sim_coax_field_res_singlexy} show FEM simulations of the electric field produced by ground (including the viewing window and slit) and the wire electrode, with a deliberately large lateral offset of \SI{0.5}{\milli\meter} in both the $x$ and $y$ directions. A schematic outline of the HV scanning head is overlaid for orientation, with the slit indicated at the bottom.
  The voltage difference between the cylinder (inner radius \SI{5.0}{\milli\meter}) and the wire (radius \SI{0.108}{\milli\meter}) is $\Delta U = -\SI{5000}{\volt}$. 
  Two-dimensional cross-sections at the $z$-midplane with \figsubref{fig:sim_coax_field_singleXY} the simulated field magnitude, and 
  \figsubref{fig:sim_coax_field_res_singlexy} the difference between the simulated realistic field with offset and the field produced by an ideal coaxial electrode system (infinitely long coaxial ground around perfectly centered wire) are shown along with an angular scale in green to be used as reference for 
  \figsubref{fig:sim_fitted_E_zTh} and \figsubref{fig:sim_fitted_res_E_zTh}. The jagged edges near the ground in \figsubref{fig:sim_coax_field_res_singlexy} are due to meshing artefacts in the FEM simulation that are not relevant for the analysis, since we focus on the appropriately densely meshed high-field region close to the wire.\\
  Subplot \figsubref{fig:sim_fitted_E_zTh} shows the electric field magnitude on a cylindrical surface (radius \SI{0.15}{\milli\meter}) around the wire, unrolled into $z$ vs. $\theta$, fitted with 
  \eqnref{sec:highres-scans:subsec:experimental-setup:eq:plateau-function} to account for wire eccentricity and boundary effects near the edges of the \SI{4}{\milli\meter} viewing window ($z=$ \SI{8}{\milli\meter} to $z=$ \SI{12}{\milli\meter}). 
  Sublot \figsubref{fig:sim_fitted_res_E_zTh} shows the difference between simulation and fit. 
  See text for more details.}
\end{figure}
At the start of each measurement, the scanning head is positioned such that the wire is centred. The high voltage is then ramped to the selected measurement value. Subsequently, the gantry automatically moves the head along the wire in \SI{4}{\milli\meter} steps, corresponding to the field of view of the high-resolution camera. At each step, an image of the wire is taken, and ten current and voltage readings are recorded by the power supply. For a given voltage difference $\Delta U$ between wire and scanning tool ground, 
the measurement results are locations on the wire vs. currents. In the following sections, one such current and voltage reading at a given location is referred to as a \textit{measurement}, a significantly high current measurement at a given location is referred to as a \textit{hotspot event}, while the location of such an event is referred to as a \textit{hotspot location} and the full measurement of all positions along a wire is called a \textit{scan}. The measurement of multiple wires between cleaning or significant change of environment is called a \textit{measurement session}. As the camera only images the wires from the top, but the scanning head applies an electric field around the entire wire, a first scan of each wire is taken with a relatively low potential of $\Delta U=-\SI{5000}{\volt}$. Afterwards, the frame is flipped and multiple measurements runs are performed with different HV settings. 
After each measurement session the frame is cleaned again in the ultrasonic bath, using deionised water at room temperature. 
This additional cleaning step is used to ensure that potential current emitting locations are due to the wire or its surface and not due to remnant dust.

\subsection{Discharge Phenomenology}
\label{sec:highres-scans:subsec:discharge-phenomenology}

The discharge phenomenology relevant for the HV wire scanning method is the dark-regime, where the discharge manifests itself only by its current, close to the transition to corona discharge, which is a form of glow discharge accompanied by visible light. 
The relevant current range extends to a few \SI{10}{\micro\ampere}. This current range is also chosen to prevent damage to the wires from prolonged corona discharge exposure in air on the timescale of a few minutes \cite{CogolloDeCadiz2021_corrosion}. A more comprehensive treatment of the relevant discharge phenomenology can be found in \textit{e.g.} \cite{raizer1997gas}. Coaxial geometries for corona discharges in air are covered extensively in the literature, see \cite{Mikropoulos:Neg_Corona_2010} for a more recent paper. In the following, \eqnref{sec:highres-scans:subsec:experimental-setup:eq:ideal-coaxial-electric-field} 
and \eqnref{sec:highres-scans:subsec:experimental-setup:eq:e-general-critical} enable us to calculate 
the “critical voltage” or “discharge inception voltage” $\Delta U_\text{c}$ for the onset of the discharge. It is defined as the voltage 
for which the surface field strength on the wire 
$\Vert E_{\text{ideal}}^{\text{coaxial}}(r) \Vert$ 
with 
$r = r_{\text{wire}}$ 
is equal to the critical field strength $\Vert E_{\text{c}}^{\text{coaxial}} \Vert$, at which a discharge current 
can be expected. For the geometry of the wire and scanning head, this value is around $-\SI{5.2}{\kilo\volt}$. The electric field of the ideal coaxial configuration is given by
\begin{align}
  \Vert E_{\text{ideal}}^{\text{coaxial}}(r) \Vert = \frac{\Delta U}{r \cdot \ln\left(\frac{R_\text{cylinder}}{r_\text{wire}}\right)} \quad ,
  \label{sec:highres-scans:subsec:experimental-setup:eq:ideal-coaxial-electric-field}
\end{align}
where 
$R_\text{cylinder}$ and $r_\text{wire}$ are the outer and inner radius. In our configuration, this is the inner radius of the copper cylinder and the wire radius, respectively \cite{raizer1997gas}. The critical field ($E_{\text{c}}$) at which the discharge ignites is given by the empirical formula, found in \cite{Mikropoulos:Neg_Corona_2010}:
\begin{align}
  \Vert E_{\text{c}}^{\text{coaxial}} \Vert \, \left[\si{\kilo\volt\per\centi\meter}\right] = A \cdot \delta \cdot \left(1 + \frac{B}{\left(\delta \cdot r_\text{wire}\, \left[\si{\centi\meter}\right]\right)^C}\right) \quad ,
  \label{sec:highres-scans:subsec:experimental-setup:eq:e-general-critical}
\end{align}
where $\delta$ is the ratio of the actual air density to the air density at $25\,\text{C}^\circ$ and \SI{1013.25}{\milli bar}. The eponymous formula by Peek \cite{peek1929dielectric} for the coaxial configuration, where the inner conductor is at a negative potential, is reproduced for the empirical factors $A=31.53$,  $B=0.305$ and $C=0.5$. Different experimental studies in air have found slightly varying values for these factors and have been noted in  \cite{Mikropoulos:Neg_Corona_2010}. 
The relative humidity of the surrounding air has a negligible effect according to the observations in \cite{peek1929dielectric}, and using the formulas in \cite{Mikropoulos:Neg_Corona_2010}.\\
{\indent}The electric field on, and close to, the wire surface for the scanning head needs to be well understood and ideally close to that of the ideal co-axial configuration. Otherwise, the measured current is  more indicative of the scanning head's specific field configuration than the intrinsic quality of the wire. The real scanning head deviates from an ideal coaxial cylinder due to its finite length and the presence of a viewing hole and slit. Furthermore, small mechanical misalignments can lead to the wire not being perfectly centred (eccentricity). Finite element method (FEM) COMSOL \cite{comsol} simulations were performed to assess the impact of such deviations.\\
{\indent}\Figref{fig:sim_coax_field_res_singlexy} shows the difference between the simulated field (derived from the FEM simulations) and the ideal coaxial field from \eqnref{sec:highres-scans:subsec:experimental-setup:eq:ideal-coaxial-electric-field}. 
For a perfectly centred wire, the simulation deviates from the ideal analytical model by \SI{0.3}{\%} due to edge effects. For eccentric wires, the field shows a dipole-like pattern with larger deviations around \SI{1}{\%} from ideal close to the wire (\figrefbra{fig:sim_coax_field_singleXY} and \figrefbra{fig:sim_coax_field_res_singlexy}). The field is enhanced on the side of the wire closest to the grounded cylinder wall.\\
{\indent}To meaningfully display the simulated three-dimensional electric field on a cylindrical surface just outside the wire, we use a empirical analytical function that captures both axial ($z$-direction) and azimuthal ($\theta$-direction) variations:
\begin{equation}
  \Vert E_{\text{fit}}(\theta, z) \Vert = \underbrace{o_p + \frac{A_p}{\left(e^{s_p(z - c_p - w_p)} + 1\right)\left(e^{s_p(-z + c_p - w_p)} + 1\right)}}_{\text{Plateau function } P(z)} + \underbrace{B_c \cos(f_c \theta - \phi_c)}_{\text{Cosine function } C(\theta)}.
\label{sec:highres-scans:subsec:experimental-setup:eq:plateau-function}
\end{equation}
The plateau component $P(z)$, formed by the product of two sigmoid functions, models the field's drop-off near the ends of the finite-length cylinder. The cosine term $C(\theta)$ accounts for the field variation around the wire's circumference caused by eccentricity. 
\Figref{fig:sim_fitted_E_zTh} and \figref{fig:sim_fitted_res_E_zTh} shows that the fit accurately reproduces the simulated field for both centred and offset cases, with small and unstructured differences, confirming the equation's validity. Another finding from \figref{sec:FEM_simulations_z05_neg05_offset_single} is, that within the central \SI{4}{\milli\meter} window visible to the \textit{Basler acA4600-7gc} camera (corresponding to one measurement step), the axial variation of the electric field is less than \SI{2}{\%} even for $\sim$\SI{0.7}{\milli\meter} (\SI{0.5}{\milli\meter} in $x$ and $y$) eccentricity wires, which is well above the \SI{0.25}{\milli\meter} centring tolerance we achieved via camera feedback. This high degree of local field uniformity is needed to ensure that each \SI{4}{\milli\meter} segment of the wire is tested under well-defined field conditions, which allows for direct comparison of measurements along the wire and justifies the application of the ideal coaxial theory as a local approximation.

\subsection{Baseline Voltage-Current Characteristics}
\label{sec:highres-scans:subsec:high-voltage-wire-scanning}

\begin{figure}
  \centering
  \includegraphics[width=0.9\columnwidth]{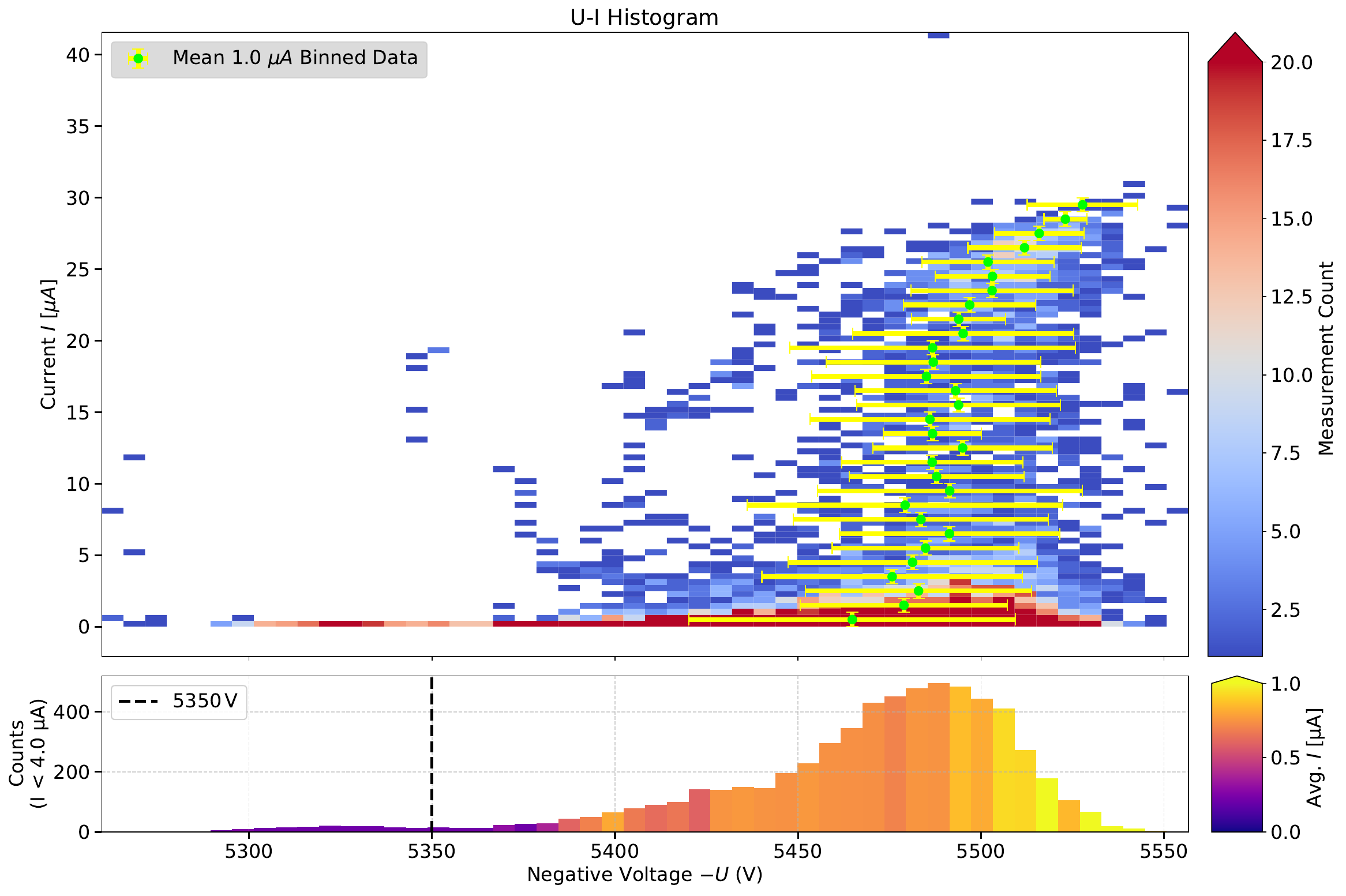}
  \caption{\label{fig::highres-scans:discharge_rampup}
    Two-dimensional histogram of voltage versus measured discharge current for the wire-ground system from 328 distinct positions for cleaned, undamaged wires. The upper panel shows the measurement density (color scale, capped at 20 counts for clarity) together with mean voltages and standard deviations for \SI{1}{\micro\ampere} current bins (green points with yellow error bars). The lower panel displays the corresponding voltage distribution for measurements with \SI{0.0}{\micro\ampere} < $I$ < \SI{4.0}{\micro\ampere}, with a dashed line at \SI{-5350}{\volt}. The adjacent colour bar represents the mean current per voltage bin, considering only values below \SI{4.0}{\micro\ampere} and truncated at \SI{1}{\micro\ampere} for readability.}
\end{figure}

Before investigating localized defects, it is essential to establish the general voltage-current ($U$-$I$) characteristic of the wire-scanner system. This baseline measurement defines the typical discharge inception voltage and current ramp-up for cleaned wires, providing a reference for identifying anomalous behaviour at specific locations. The measured relationship between the applied voltage and the resulting current during the onset of the dark discharge is detailed in \figref{fig::highres-scans:discharge_rampup} with a two-dimensional histogram of the resulting $U$-$I$ data, where the colour scale indicates the measurement frequency at each point in the $U$-$I$ space. Shown in the plot are measurement sessions of multiple new \SI{880}{\milli\meter} long wires. To ensure reliability, the data has been subjected to quality cuts. For example, incomplete measurement scans and datasets exhibiting transient, unstable high-current spikes at low voltages that did not persist as the voltage increased, were cut. Additionally, during ramp-up a threshold current of \SI{25}{\micro\ampere} was set , leading to a slight increase in the number of observed currents right above this region. The inherent stochasticity of the discharge inception process is visible from the spread in the data, particularly at lower currents.\\
{\indent}To quantify the average behaviour, several statistical representations are overlaid on the histogram. The green points with horizontal yellow error bars show the mean voltage and its standard deviation, respectively, calculated for discrete \SI{1}{\micro\ampere} current bins, highlighting the most probable $U$-$I$ characteristics during operation characterized by a lack of dark current below a critical threshold voltage before a sharp rise in current in the order of \SI{10}{\micro\ampere}.\\
{\indent}Prior to data collection, all wires were subjected to the standardized cleaning procedure detailed in \secref{sec:introduction} and \secref{sec:highres-scans:subsec:setup-measurement-procedure}. However, only half of the wires were treated with a citric acid passivation process, while the other half were not. No statistically significant difference was observed in the $U$-$I$ characteristics between these two sample sets; therefore, the data from both are combined in this analysis. This is in slight contrast to the study done by Tomás \textit{et al.} \cite{TOMAS201849} in which nitric acid cleaning and passivation leads to a significant decrease of electron emission activity. However, their method and single-electron sensitivity deviated significantly from our work. By extrapolating the trend of the mean $U$-$I$ data points to zero current, the discharge inception voltage $\Delta U_\text{c}$ can be estimated. From \figref{fig::highres-scans:discharge_rampup} one can see that the experimental inception voltage is observed to be approximately $-\SI{5460 \pm 50}{\volt}$ judging by mean and standard deviation of the 0-\SI{1}{\micro\ampere} bin voltage. This value is in good agreement with the theoretical prediction of around $-\SI{5200}{\volt}$, derived from empirical models for negative corona discharges in this specific geometry~\cite{Mikropoulos:Neg_Corona_2010}, \textit{cf}. \secref{sec:highres-scans:subsec:discharge-phenomenology}. The measured value falls within \SI{5}{\%} of the theoretical prediction, a deviation from experiment also acknowledged by the authors \cite{Mikropoulos:Neg_Corona_2010}. \Figref{fig::highres-scans:discharge_rampup} also includes a subplot showing the voltage distribution corresponding to measured currents below \SI{4.0}{\micro\ampere}. This analysis helps identify a suitable operating voltage for scanning that remains outside the high-current regime of clean, undamaged wires. A value of \SI{-5350}{\volt} satisfies this criterion, and the accompanying color scale indicates that currents at and below this voltage remain under \SI{0.3}{\micro\ampere}, which can therefore be taken as a representative background level.

\subsection{Investigation of Natural Discharge Hotspot Locations}
\label{sec:highres-scans:subsec:visual_correlation}

With the baseline electrical behaviour established, we investigated naturally occurring discharge hotspot events on pristine wires. The primary goals are to assess discharge repeatability and to determine if they correlate with static visual features that could be identified via optical inspection. 
The key objective is to assess whether (\textit{i}) visual inspection, (\textit{ii}) discharge current under HV stress, or (\textit{iii}) both together, can screen for discharge-prone wire regions. We therefore compared wire images (\figrefbra{fig:experimental_setup_camera_image}) with simultaneously measured currents, following the procedure in \secref{sec:highres-scans:subsec:setup-measurement-procedure}.

\subsubsection{Transient Nature and Repeatability of Hotspot Locations}
\label{sec:highres-scans:subsec:repeatability_and_cleaning_effect}
\begin{figure}
  \centering
  \includegraphics[width=.99\columnwidth,trim= 0 0 0 40, clip=true]{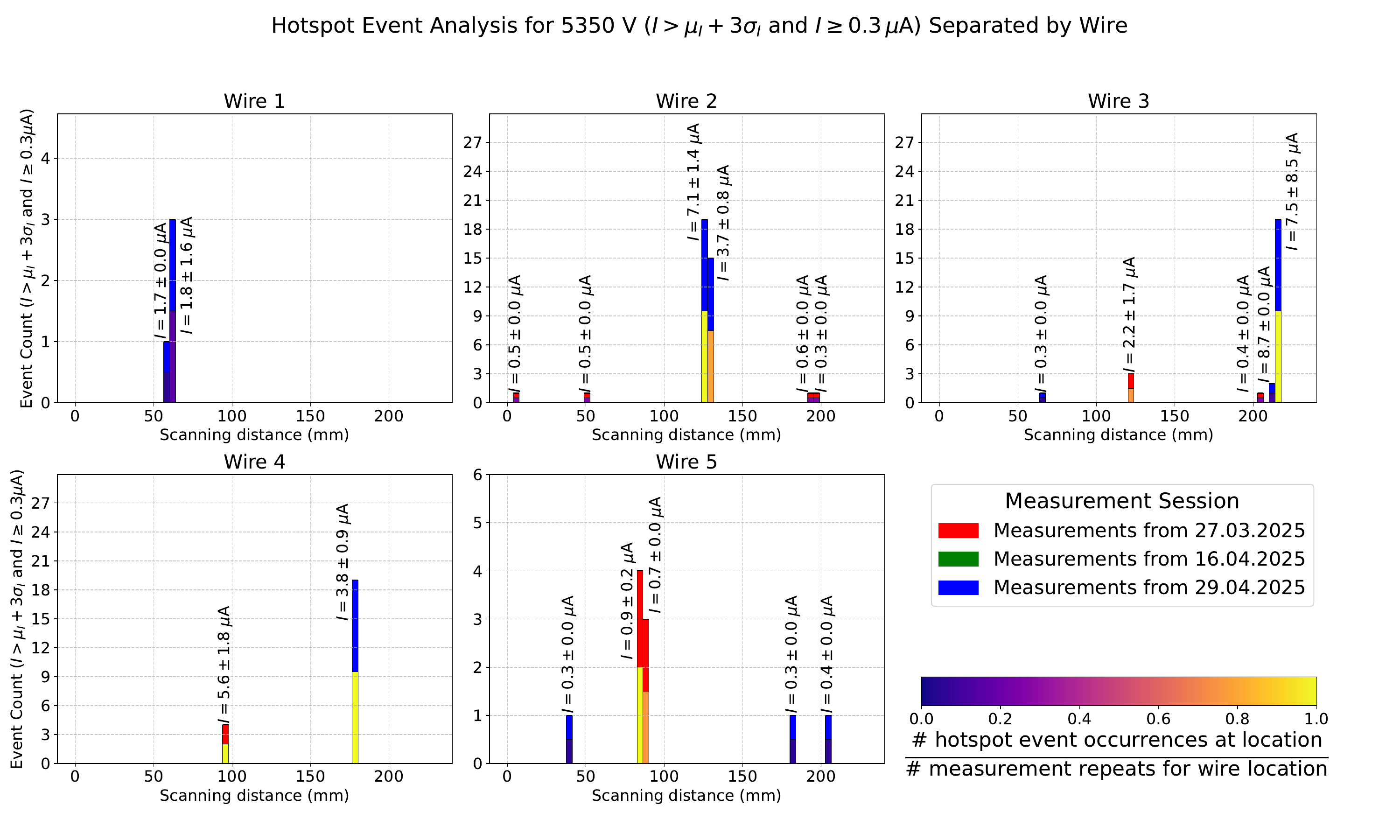}
  \caption{\label{fig::highres-scans:repeatability_cleaning_effect}Analysis of hotspot location and consistency for five test wires. The histograms show the number of significant current events ($I_\text{e} > \mu_{I_\text{r}} + 3\,\sigma_{I_\text{r}}$ and $I_\text{e} \geq \SI{0.3}{\micro\ampere}$ for a voltage of $-\SI{5350}{\volt}$) detected at locations \SI{4}{\milli\meter} apart along each wire. The colour of the top half of each bar indicates the measurement session, demonstrating that hotspot events appear/disappear after cleaning. The colour of the bottom half represents the hotspot event consistency for repeated scans in a single measurement session (a score of 1.0, yellow, signifies the hotspot event occurred at that location in every scan of a measurement session), demonstrating high short-term repeatability.}
\end{figure}

The first step in understanding the hotspot locations was to determine if they were persistent phenomena tied to the wire's permanent structure. We therefore analysed the location and consistency of high-current events across multiple scans and cleaning cycles. 
The results of this study are summarized in \figref{fig::highres-scans:repeatability_cleaning_effect}. 
The figure displays histograms of the hotspot event count for five individual wires at a constant voltage of $-\SI{5350}{\volt}$. 
A measurement is classified as a hotspot event when the measured current $I_\text{e}$ exceeds the wire's mean current over a scan $\mu_{I_\text{r}}$ by more than three standard deviations ($I_\text{e} > \mu_{I_\text{r}} + 3\,\sigma_{I_\text{r}}$), with a minimum threshold of $I_\text{e} \geq \SI{0.3}{\micro\ampere}$. The analysis reveals two key observations:\\
{\indent}First, there is a high degree of repeatability for hotspot locations within a measurement session of measurement scans, between the repeated cleaning of the frame. The repeatability is represented by the colour of the bottom half of each histogram bar. Each measurement of the first day was repeated 4 times, while measurements on the second and third day were repeated 19 times. The score is the ratio of hotspot events over the total number of measurements at that specific \SI{4}{\milli\meter} scan location within a measurement session. 
Several prominent hotspot locations across all five wires exhibit high consistency scores. Notably, regions exhibiting elevated average currents are often adjacent to positions with slightly lower average currents and reduced consistency scores. This behaviour can be attributed to the approximately 3\% decrease in the electric field outside the \SI{4}{\milli\meter} scan window (see the bottom and top edges of \figrefbra{fig:sim_fitted_E_zTh}), which suppresses discharge activity to some extent and leads to a more intermittent appearance.
Such repeatability within a single measurement session speaks for the reliability of our methodology.\\
{\indent}The second key observation concerns the effect of the cleaning procedure on hotspot locations. The top half of each bar is coloured according to the measurement session in which the hotspot events occurred. Each distinct hotspot location is exclusively associated with a single measurement session (\textit{i.e.}, a single colour). No instance was observed where hotspot locations from different measurement sessions, separated by a cleaning cycle, contributed to a hotspot event at the same location. This demonstrates that the ultrasonic cleaning process effectively removes or alters the surface features responsible for the hotspot events.\\
{\indent}Taken together, these findings suggest that the majority of the observed hotspot events for cleaned, pristine wires are not caused by permanent macroscopic defects or damage to the wire surface. Instead, they appear to originate from transient surface features, such as microscopic contaminants or asperities, which are stable enough to produce consistent discharges over multiple scans but are removed by the cleaning protocol. 

\subsubsection{Microscopic Feature Analysis with an Autoencoder}
\label{sec:highres-scans:subsec:AE_feature_analysis}

\begin{figure}
\hspace{-0.08\textwidth}
    \subfloat[]{
    \label{fig:highres-scans:naive_autoencoder_reconstruction_loss_sig_back_current}
    \includegraphics[height=0.23\textheight]{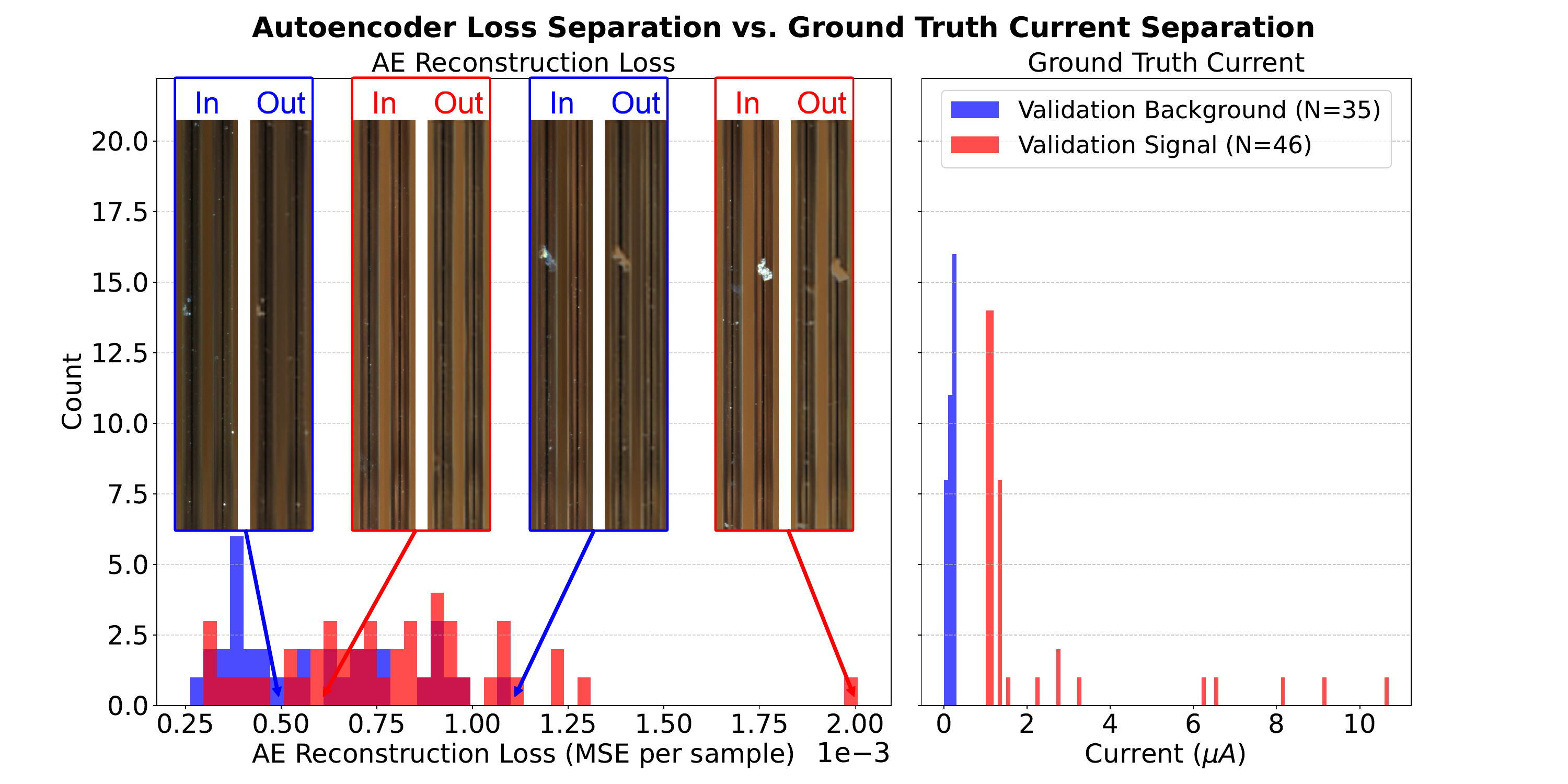}}
    \hspace{-0.08\textwidth}
  \subfloat[]{
    \label{fig:highres-scans:naive_autoencoder_reconstruction_loss_all_destroyed_wires}
    \includegraphics[height=0.23\textheight]{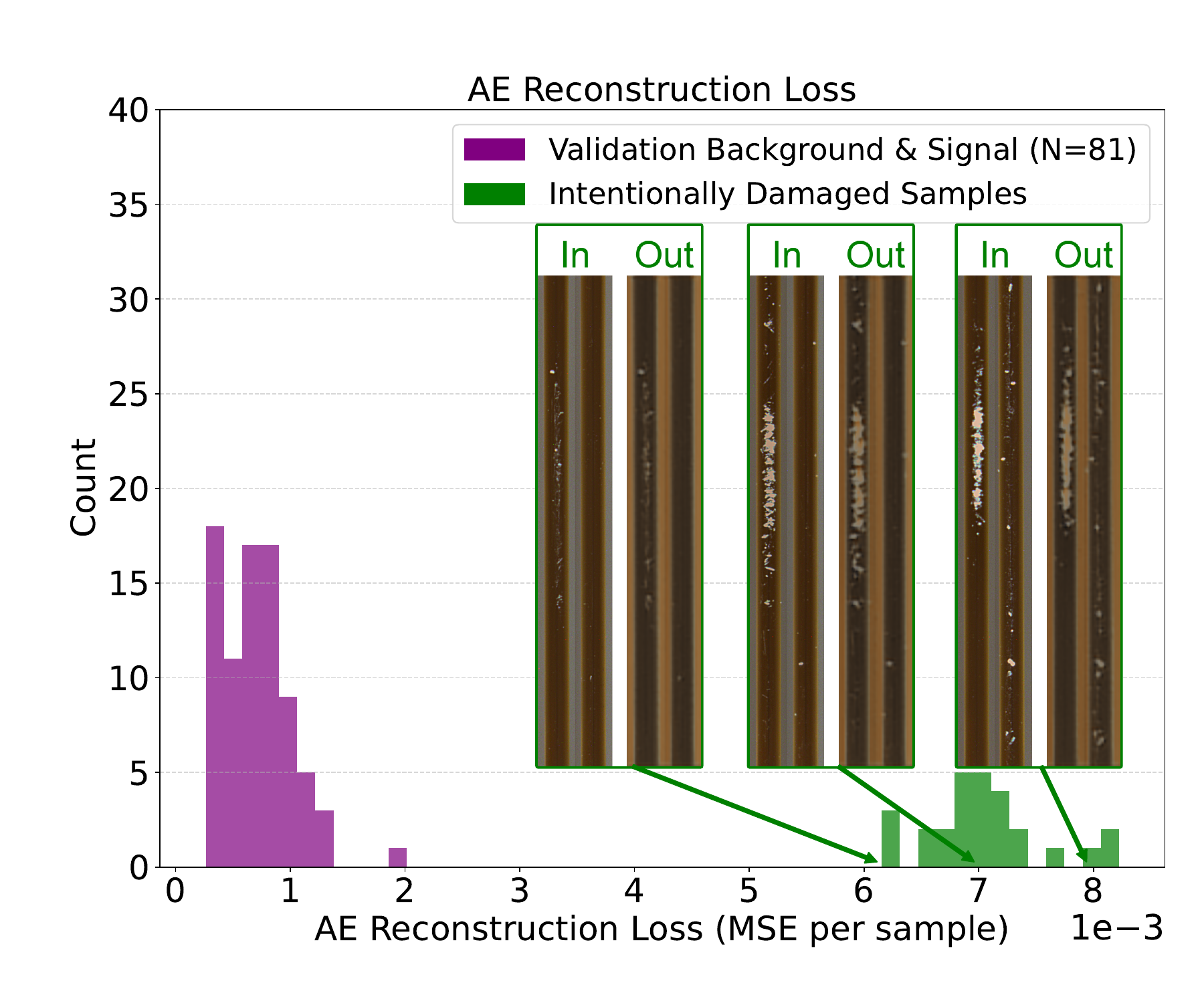}}
  \caption{\label{fig::highres-scans:naive_autoencoder_reconstruction_loss}MSE losses for an autoencoder trained on wire images with low-current discharges ($I<\SI{0.3}{\micro\ampere}$) at \SI{-5350}{\volt}. \figsubref{fig:highres-scans:naive_autoencoder_reconstruction_loss_sig_back_current}: Reconstruction loss and current distributions for low-current (``background'',  blue) and higher-current ($I>\SI{1}{\micro\ampere}$, ``signal'',  red) samples. Overlapping distributions indicate no clear visual correlation between wire features and discharge activity. \figsubref{fig:highres-scans:naive_autoencoder_reconstruction_loss_all_destroyed_wires}: Losses for all samples (purple) vs.\ intentionally damaged wires (green) showing clear separation; such wires are utilized in \secref{subsec:highres-scans:subsec:Induced_Surface_Damage}. To illustrate the corresponding defects on wires, wire images are added to both figures.  Each ``In'' image is a stitched front- and backside image of a wire, which is fed to the input of the autoencoder, whilst ``Out'' images are the corresponding encoded and decoded images.}
\end{figure}
Having established that natural hotspot locations are transient, we investigated if the visual condition of the surface of a wire (top and bottom surfaces, the top surface is seen in \figref{fig:experimental_setup_camera_image}), could be used to predict hotspot locations. To test this hypothesis in a quantitative and unbiased manner, an unsupervised machine learning approach using an autoencoder is employed, a technique often utilized for anomaly detection \cite{BELIS2024100091}.\\
{\indent}A convolutional autoencoder was trained on stitched front- and backside images of the wire, each with a resolution of $1600\times\SI{240}{px}$ (height $\times$ width), representing the full visible region of the wire surface. Only images from regions with low current activity ($I < \SI{0.3}{\micro\ampere}$ at $\SI{-5350}{\volt}$) were used for training, allowing the model to learn the reconstruction of nominal, undisturbed wire surfaces. The network architecture consisted of convolutional layers with a kernel size of 3 and filter depths of $\{16, 8, 4\}$, without batch normalization. Training was performed using the Adam optimizer (learning rate = 0.001) for 100 epochs with a batch size of 8.\\
{\indent}After training, the model was evaluated on three distinct, previously unseen datasets:
\begin{itemize}
    \item \textbf{Background:} nominal wire images showing low current activity  ($I<\SI{0.3}{\micro\ampere}$ at $-\SI{5350}{\volt}$),
    \item \textbf{Signal:} wire regions exhibiting elevated current activity ($I>\SI{1}{\micro\ampere}$ at $-\SI{5350}{\volt}$),
    \item \textbf{Damaged:} wire sections with intentionally introduced surface defects (\textit{cf}. \secref{subsec:highres-scans:subsec:Induced_Surface_Damage}) .
\end{itemize}
If the signal images contained visual features associated with hotspot locations, the model, trained exclusively on background images, would be expected to reconstruct them less accurately, resulting in a higher reconstruction loss (\textit{i.e.} mean squared error, MSE).  This is in contrast to \cite{Deisting2025}, where only image data was available. The distribution of reconstruction losses for the background and signal datasets is shown in \figref{fig:highres-scans:naive_autoencoder_reconstruction_loss_sig_back_current}. The histogram for the signal data (red) overlaps almost entirely with that of the background data (blue), showing no clear separation or distinct high-loss region, even though their underlying current distributions differ. This indicates that the autoencoder did not identify any consistent visual differences between low-current and high-current regions.\\
{\indent}The efficacy of the trained model is, however, demonstrated by its response to the intentionally damaged wires, as shown in \figref{fig:highres-scans:naive_autoencoder_reconstruction_loss_all_destroyed_wires}. In this case, the reconstruction loss is clearly higher than for nominal wire regions, confirming that the model is capable of distinguishing significant, visually apparent surface anomalies. For regular wire surfaces, however, the absence of a measurable separation between signal and background regions suggests that persistent hotspot locations cannot be reliably discerned from image data alone. This underlines the necessity of the employed wire scanning approach, which combines spatially resolved imaging with localized current measurements to characterize discharge-prone regions.

\subsubsection{Image Blurriness as a Macroscopic Indicator}

\begin{figure}
  \centering
  \includegraphics[width=1.0\columnwidth]{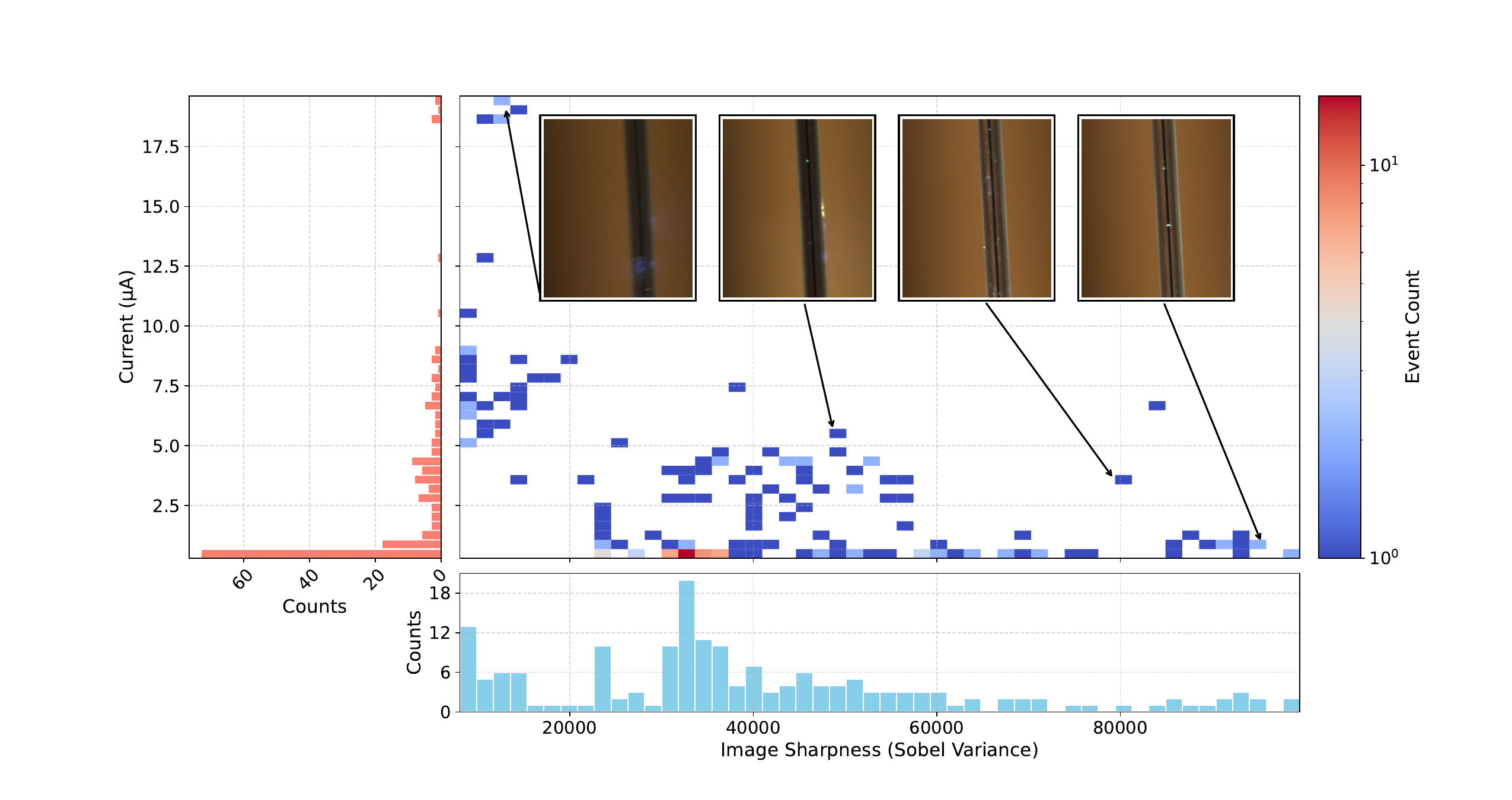}
  \caption{\label{fig::highres-scans:blurriness_study}Two-dimensional histogram of image sharpness (Sobel variance) versus measured current at \SI{-5350}{\volt}. Higher Sobel variance corresponds to a sharper image. The distribution shows an anti-correlation, with hotspot events exclusively occurring when the wire image is blurry. The inlaid images exemplify this relationship, showing a blurry image for high current and sharper images for lower currents.}
\end{figure}

During the measurements, it was qualitatively observed that the live image of the wire often appeared blurred whenever the power supply indicated a high current, even in the absence of a visibly glowing corona discharge. Reason for the blurriness are vibrations in combination with an effectively long exposure time of \SI{1}{\second}. The vibrations are hypothesized to be induced by the ionic wind generated during the discharge process \cite{ionic_wind_corona_2010}.
The converse, namely that an external driving force responsible for vibrations is causing the increased current, is much less likely since the vibration amplitudes observed are $\leq$\SI{100}{\micro\meter} cannot account for a significant change in discharge current according to the FEM analysis in \secref{sec:highres-scans:subsec:discharge-phenomenology}.\\
{\indent}\Figref{fig::highres-scans:blurriness_study} presents a two-dimensional histogram of the image sharpness versus the measured current. The image sharpness is calculated using the variance of the Sobel operator (applied on the centre 30\% of the wire surface), a common metric in image processing where higher values correspond to sharper images \cite{sobel_edge_2014}. The data populates a distinct region in the lower-left portion of the plot. The top-right region is devoid of data, indicating that no instances of high current were recorded when the wire image was sharp. The inlaid images exemplify this. 
This negative correlation suggests that image blurriness can serve as a rapid, real-time qualitative indicator of significant discharge activity. While not a substitute for precise localization, this visual cue could be used in future automated systems to trigger more detailed scans.

\subsection{Investigating Induced Abrasive Surface Damage}
\label{subsec:highres-scans:subsec:Induced_Surface_Damage}

Most natural discharge hotspot locations are transient without clear visual signatures. To link abrasive damage, which could plausibly occur when mounting and unwinding the wires, with discharge behaviour, we tested wires with induced abrasive damages, enabling us to determine the sensitivity of our scanning method to given surface morphologies.

\subsubsection{Experimental Procedure}

Pristine wires were first cleaned with isopropyl alcohol and lint-free optical wipes, then their baseline electrical performance was characterised (as exemplified in \figrefbra{fig::highres-scans:discharge_rampup}) at discrete locations (\SI{1}{\centi\meter} apart) and imaged with the confocal microscope referenced to at the beginning of \secref{sec:setup}. Because of reflectivity and curvature, only a $\sim$20–30° surface arc was resolvable. Abrasive damage was then applied using sub-millimetre strips of sandpaper, with one or two transverse strokes localized to the microscope’s viewing area. After abrasion, the sites were cleaned again with alcohol and optical wipes to remove loose debris, re-imaged, and rescanned. The untouched backside of the wire was also imaged to verify that no unintended damage had occurred. For visual examples of the front and back-side of typical damaged wire samples see the autoencoder input samples in \figref{fig:highres-scans:naive_autoencoder_reconstruction_loss_all_destroyed_wires}.

\subsubsection{Correlation between Surface Roughness and Discharge Voltage}
\label{subsubsec:highres-scans:subsec:Corr_Roughness_Voltage}
\begin{figure}
    \subfloat[]{
    \label{fig::highres-scans:surface_scan_example}
    \includegraphics[width=0.85\linewidth]{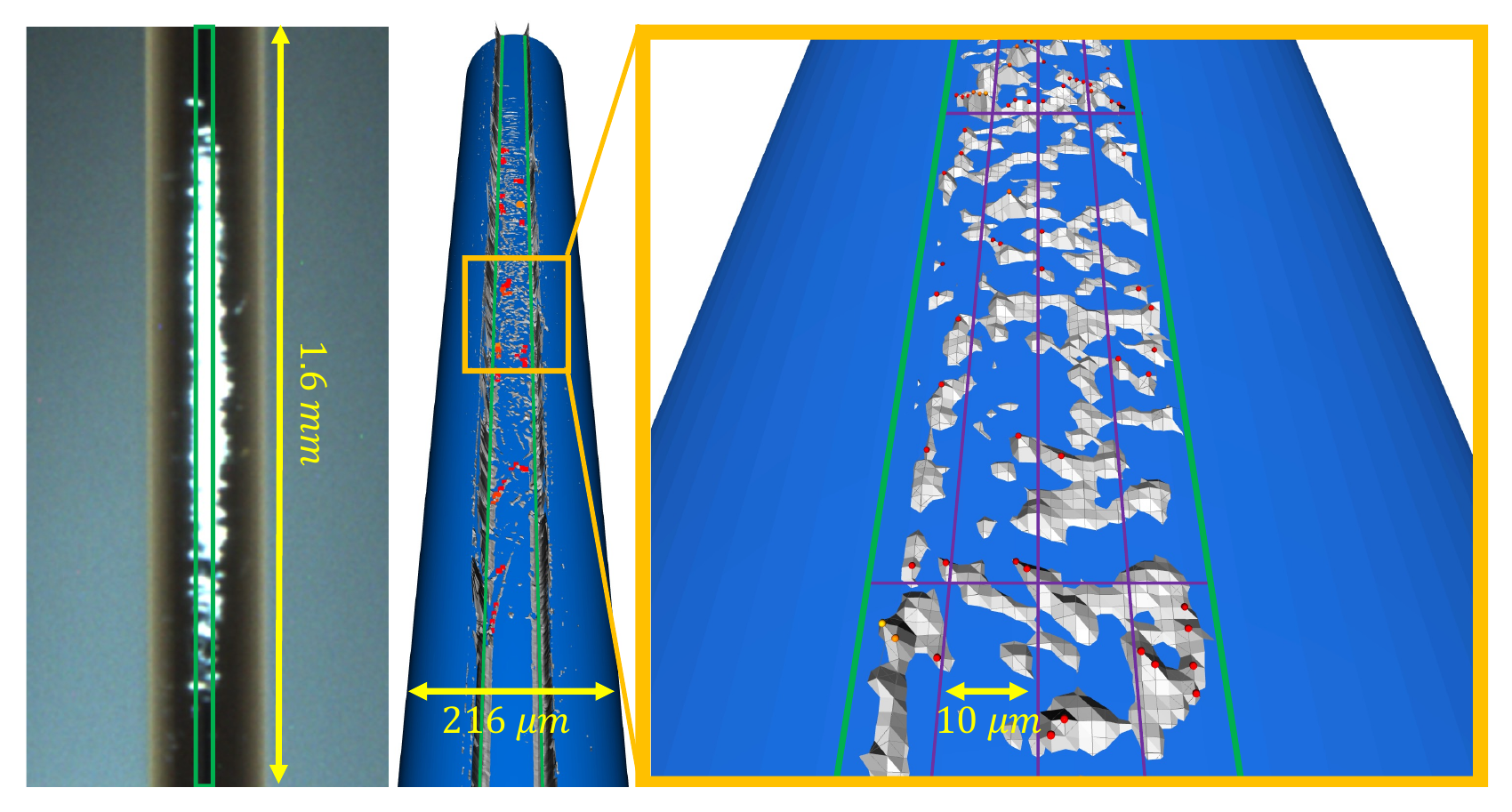}}
  \subfloat[]{
    \label{fig::highres-scans:roughness_counts_vs_voltage}
    \includegraphics[width=0.90\linewidth]{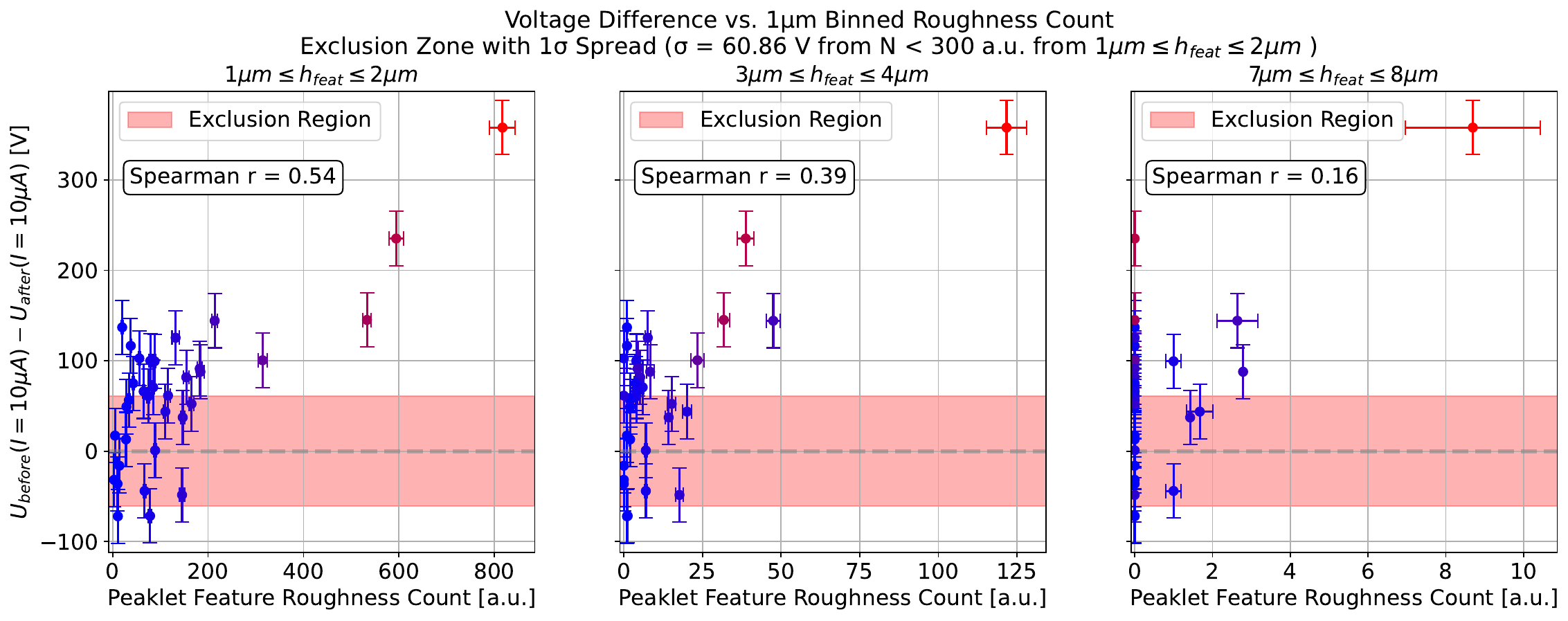}}
  \caption{\label{fig::highres-scans:roughness_counts_vs_voltage_with_example}\figsubref{fig::highres-scans:surface_scan_example}: Example surface scan of an intentionally damaged wire segment.
    (Left) Optical image acquired through a telecentric lens, with the effective usable field of view (FOV) of the confocal microscope indicated by the green box (\SI{1.6}{\milli\meter} in length).
    (Center) Reconstructed 3D wire surface with a fitted blue cylindrical model of radius \SI{0.108}{\milli\meter}. The green lines denote the microscope’s FOV, while red and orange markers indicate the final selection of protrusion peaks.
    (Right) Magnified view of the central region after removing reflection artefacts and aligning the surface with the fitted cylinder. Shown here is the ungrouped result of the peak-finding algorithm, displaying all local maxima with a $k=4$ graph-based approach. Further selection was done by filtering out peaks that are local maxima compared to their surrounding by less than \SI{0.1}{\micro \meter} (potentially flat regions within microscopes sensitivity). The purple grid divides the FOV into the 24 subregions used for the local error estimation.
    \figsubref{fig::highres-scans:roughness_counts_vs_voltage}: The number of surface peaklets of a given height, counted via confocal microscopy, versus the reduction in voltage needed to achieve a \SI{10}{\micro\ampere} current after abrasion. Subplots correspond to feature heights of 1-\SI{2}{\micro\meter}, 3-\SI{4}{\micro\meter}, and 7-\SI{8}{\micro\meter}. The data points are coloured according to their horizontal axis order in the first subplot on the left, to alleviate tracking the points across subplots. As each point corresponds to a different wire segment measured, the colour encodes the wire segment across the three height categories. The positive correlation shows that increased surface roughness lowers the required discharge voltage. The red band indicates the $1\,\sigma$ noise level, with points outside representing significant damage.}
\end{figure}
{\indent}After applying corrections for the wire's curvature, the metric used to quantify the induced damage to the wire surface was defined as the total count of microscopic surface peaks (``peaklets'') within specific height ranges. A significant challenge in this analysis is the limited viewing angle of the microscope. To estimate the total peaklet count over the entire damaged area (which was determined from macroscopic top-view photographs), the count from the visible region was extrapolated. The uncertainty in this extrapolation was estimated by subdividing the visible area into 24 smaller regions, calculating the standard deviation of the peaklet count density across these regions, and propagating this variance to the total extrapolated count (\textit{cf}. \figrefbra{fig::highres-scans:surface_scan_example}).\\
{\indent}The primary results of this study are the three panels of \figref{fig::highres-scans:roughness_counts_vs_voltage}. They show the correlation between the peaklet count for three different feature height ranges (1-\SI{2}{\micro\meter}, 3-\SI{4}{\micro\meter}, and 7-\SI{8}{\micro\meter}) and the change in voltage required to produce a \SI{10}{\micro\ampere} current. This current value was chosen as it represents a stable discharge well above background activity (\textit{cf}. \figrefbra{fig::highres-scans:discharge_rampup}), providing a reproducible metric for comparison. The voltage change is calculated as the voltage difference before minus after abrasion ($U_{\text{before}} - U_{\text{after}}$). A positive value signifies that a lower voltage is needed to achieve the same current after damaging the wire. The error on the voltage difference is estimated to be \SI{30}{\volt}, corresponding to a positional uncertainty of \SI{0.25}{\milli\meter} (wire-ground distance) between consecutive scans, based on the electrostatic simulations discussed in \ref{sec:highres-scans:subsec:discharge-phenomenology} (\textit{e.g.} \figrefbra{sec:FEM_simulations_z05_neg05_offset_single}).\\
{\indent}The plots reveal a positive correlation: an increase in the number of surface features corresponds to a larger reduction in the voltage required for discharge, as quantified by the positive Spearman correlation coefficients \cite{Spearman_Original_1904}. This trend is present across all feature sizes, though the correlation appears strongest for the most numerous 1-\SI{2}{\micro\meter}. 
Such a decrease in inception voltage depending on the presence of surface asperities has been noted by others especially for positive DC coronas \cite{Bian_AC_corona_discharge:2011, Bian_Surface_Roughness_Corona:2012, Surface_Morph_Pos_Cor:2019, TiO2_coating_against_surface_discharge:2022}. The red band in \figref{fig::highres-scans:roughness_counts_vs_voltage} represents the $1\,\sigma$ spread of $U_{\text{before}} - U_{\text{after}}$ values around \SI{0}{\volt} in samples with low roughness counts ($<300$) in the 1-\SI{2}{\micro\meter} data, which has the highest abundance of non-zero count values. This band is an exclusion zone to distinguish significant effects from measurement noise and the inherent stochasticity of the discharge process (\textit{cf}. \figref{fig::highres-scans:discharge_rampup}), \textit{i.e.} data points lying within this band cannot be given statistical significance. The results suggest that under the present geometry, currents in the order of a few microamperes for a measurement voltage reduced more than \SI{200}{\volt} below what is necessary for typical undamaged wires to exhibit the same discharge current, is a robust indicator of significant abrasive damage. This value could serve as a practical threshold in a quality control context for identifying compromised wires.

\section{Conclusions and Outlook}
\label{sec:conclusions:outlook}

To meet the stringent requirements for large-area electrodes in future dual-phase TPCs, we have developed and characterized a versatile, high-resolution scanning platform for comprehensive quality assurance.\\
{\indent}We extended the GRANITE platform by developing a coaxial scanning head for localized non-contact, and non-destructive high-voltage qualification. We found that naturally occurring discharge hotspot events are transient and not well-correlated with static visual features, rendering simple optical inspection insufficient for electrical screening. These type of discharge hotspot events depend on wire preparation, in the sense that they are removed by re-cleaning. We established a quantitative link between induced abrasive surface damage and a reduction in discharge inception voltage, validating the scanner's sensitivity to physical defects. Only these intentional created hotspots are not transient. A voltage reduction of $\sim\!\!\SI{200}{\volt}$ serves as a robust indicator of compromising surface asperities.\\
{\indent}In summary, this work provides a powerful toolkit combining precise mechanical metrology with localized electrical screening, enabling a systematic, data-driven approach to electrode characterization for future low-background experiments. A planned upgrade using a collaborative robotic arm will extend these capabilities to the \SI{3}{\meter}-scale electrodes required for next-generation dark matter observatories.

\section*{Acknowledgements}

This work has been supported by the Cluster of Excellence “Precision Physics, Fundamental Interactions, and Structure of Matter” (PRISMA$^{+}$ EXC 2118/1) funded by the German Research Foundation (DFG) within the German Excellence Strategy (Project ID 390831469). Further support was received by BMBF under grant number 05A20UM1. D. Wenz was supported by the German Academic Scholarship Foundation. The authors thank L. F. Deibert for her help during the data taking.


\bibliographystyle{JHEP}
\bibliography{notes.bib}

\end{document}